\documentclass[twocolumn,superscriptaddress]{revtex4-2}
\usepackage{amsmath, amsfonts, amssymb, graphicx, color,bm}
\usepackage{xcolor}

\newcommand{\bphi}{{\bm\phi}}
\newcommand{\beq}{\begin{equation}}
\newcommand{\eeq}{\end{equation}}
\newcommand{\bea}{\begin{eqnarray}}
\newcommand{\eea}{\end{eqnarray}}

\newcommand{\seeSI}{SI Text}

\graphicspath{{./figures/}}

\newcommand{\add}[1]{{#1}}

\usepackage{url}

\newcommand{\lastequal}{Corresponding authors. These authors contributed equally.}

\begin{document}

\title{\add{Theoretical limits for sensing through phase separation}}

\author{Henry Alston}
\affiliation{Laboratoire de Physique, \'Ecole Normale Sup\'erieure, CNRS, PSL Universit\'e, Sorbonne Universit\'e,
Universit\'e de Paris, 75005 Paris, France}
\author{Mason Rouches}
\affiliation{The James Franck Institute, The University of Chicago, Chicago, Illinois 60637, USA}
\affiliation{Department of Physics, The University of Chicago, Chicago, Illinois 60637, USA}
\author{Arvind Murugan}
\thanks{\lastequal}
\affiliation{The James Franck Institute, The University of Chicago, Chicago, Illinois 60637, USA}
\affiliation{Department of Physics, The University of Chicago, Chicago, Illinois 60637, USA}
\author{Aleksandra M. Walczak}
\thanks{\lastequal}
\affiliation{Laboratoire de Physique, \'Ecole Normale Sup\'erieure, CNRS, PSL Universit\'e, Sorbonne Universit\'e,
Universit\'e de Paris, 75005 Paris, France}
\affiliation{The James Franck Institute, The University of Chicago, Chicago, Illinois 60637, USA}
\affiliation{Department of Physics, The University of Chicago, Chicago, Illinois 60637, USA}
\author{Thierry Mora}
\thanks{\lastequal}
\affiliation{Laboratoire de Physique, \'Ecole Normale Sup\'erieure, CNRS, PSL Universit\'e, Sorbonne Universit\'e,
Universit\'e de Paris, 75005 Paris, France}
\affiliation{The James Franck Institute, The University of Chicago, Chicago, Illinois 60637, USA}
\affiliation{Department of Physics, The University of Chicago, Chicago, Illinois 60637, USA}

\begin{abstract}
Biomolecular condensates form on timescales of seconds in cells upon environmental or compositional changes. Condensate formation is thus argued to act as a mechanism for sensing such changes and quickly initiating downstream processes, such as forming stress granules in response to heat stress and amplifying cGAS enzymatic activity upon detection of cytosolic DNA.  \add{Here, we study a dynamical model of droplet nucleation and growth to demonstrate how phase separation allows cells to discriminate small concentration differences on finite, biologically relevant timescales.} We propose optimal sensing protocols, which use the sharp onset of phase separation. We show how, given experimentally measured rates, cells can achieve rapid and robust sensing of concentration differences of $1\%$ on a timescale of minutes, offering an alternative to classical biochemical mechanisms.

\end{abstract}

\maketitle

\section{Introduction}

Biomolecular condensates formed through liquid-liquid phase separation provide membraneless compartmentalisation in the cell \cite{Hyman2014,Banani2017}. While the physical principles governing these condensates  \cite{Jacobs2013, Riback2020, Jacobs2021} and the role of non-equilibrium processes \cite{Zwicker2015, Weber2019, Folkmann2021} remain under much scrutiny, they allow for the localisation of biochemical processes 
finding broad functionality \cite{Zwicker2014, Su2016, Hnisz2017,Wei2020, Klosin2020-ca} including ribosome production in the nucleolus \cite{Lafontaine2021} and establishing polarity in asymmetric cell division \cite{Brangwynne2009, Lee2013}. 

Unlike membrane-bound organelles, condensates can form and dissolve rapidly, potentially allowing for fast responses to changing conditions. This makes phase separation particularly well suited for formulating stress responses in the cell \cite{Riback2017}, such as forming stress granules under heat shock \cite{Wallace2015}, terminating translation during starvation \cite{Franzmann2018} or creating foci in response to DNA damage \cite{Heltberg2022}. Its reversible nature provides cells with a mechanism to sense and respond to small fluctuations in internal and environmental signals \cite{Yoo2019}. Changes in composition can also trigger phase separation, endowing cells with an ability to sense whether the concentrations of signalling molecules are above a threshold set by the transition point to a phase-separated state. Cytosplasmic sensors are argued to exploit this mechanism to regulate the amount of cytosolic DNA in the cell \cite{Du2018}.

\add{In these examples, the formation of droplets signifies both the sensing of a change in signal and the initiation of the physical response, potentially without the need for a separate downstream process. Understanding the limits of phase separation as a sensor would clarify its role across this broad range of processes.} Perhaps the simplest task asked of a sensing mechanism is to measure the abundance or concentration of signalling molecules. Fundamental limits for the case of ligand-receptor binding systems were derived in the seminal work of Berg and Purcell \cite{Berg1977, Bialek2005} but similar problems find much recent interest in chemotaxis and development~\cite{Endres2006, Endres2009, Mora2010, Mora2015, Mora2019b, Desponds2020}. In this work, we ask under what circumstances can the process of phase separation most effectively sense or discriminate concentrations in simple multicomponent fluid mixtures. By phase separating (or not), droplets implicitly reflect some measurement of concentration relative to a pre-determined value: we quantify here how fast and robust this measurement is and thus how reliable condensate formation is as a trigger for downstream processes. 

In practice, cells use a variety of sensing mechanisms. Structural cooperativity such as allosteric regulation \cite{Monod1965} is well understood to provide sensitive responses to molecular signals. Goldbeter-Koshland kinetics describe how small differences in the activity of antagonistic enzymes can drive large fluctuations in the relative abundance of two substrate forms \cite{Goldbeter1981, Goldbeter1984}. The phosphorylation of Cdk1 functions as a regulatory switch that enables ultrasensitive responses in Cdc25C activation, occurring on timescales as short as 30 minutes \cite{Trunnell2011}. A phase transition to a phase-separated state naturally offers a sharp, threshold-like response: an arbitrarily small change in e.g.\,concentrations leads to very different equilibrium (long-time) states \cite{Bray2002}. {Phase transitions for liquid \cite{Braz-Teixeira2024-jl,Chalk2024-em} and solid or crystalline \cite{Murugan2015-ps,Zhong2017-kc, Evans2024-se} phases have been proposed as sensors of concentrations in multicomponent molecular mixtures.
However, achieving sharp transitions in these mechanisms takes time. Employing phase separation as a \textit{finite-time} sensing mechanism would actually rely on the non-equilibrium dynamics, 
begging the question as to how it performs against these other sensing mechanisms on realistic biological timescales. 

\add{Building on classical theory for phase separation in fluid mixtures \cite{Huggins1941, Flory1941, Cahn1958, Oxtoby1992}, which is now well analyzed in the context of biomolecular condensates \cite{Zwicker2015,Mao2019,Weber2019,Shimobayashi2021}, we derive a dynamical description for the nucleation and growth of droplets in a ternary fluid. Properties of ternary phase separating mixtures have received much attention, including spatial Cahn-Hilliard models derived from explicit lattice models \cite{Hoyt1989}, extensions of classical nucleation theory to describe nucleation near the spinodal concentrations \cite{Unger1984}, characterizations of the critical nucleation seed \cite{Philippe2014, Philippe2015, Philippe2016} and coarsening dynamics in the presence of off-diagonal terms in the diffusion tensor \cite{Philippe2013}. Our work unifies droplet nucleation and growth into a functional model of ternary phase separation to probe the timescales involved with forming droplets.} Our choice of a ternary mixture will allow us to have different subsystems interact with each other through common molecules. \add{We distinguish functional molecules $A$, confined to individual subsystems, from scaffold molecules $S$ that are required in both systems for forming droplets.}
Implicitly, we assume that these molecules are at moderate concentrations, a necessity for phase separation to be realized.

Using this most general model,
we first consider the problem of sensing whether a concentration of molecules $A$ is above or below the phase transition. We identify fundamental limits to how accurately this can be inferred from droplet formation in finite time. We also compare these limits to other classical sensing mechanisms such as ligand-receptor binding. The role of a second phase separating system with a fixed concentration of functional molecules (effectively acting as a measuring stick) is then discussed. We show how competition for limiting shared resources between the mixtures enhances sensitivity before identifying a minimal set-up which optimally employs phase separation for distinguishing concentrations. {Under such conditions, we demonstrate that concentration differences of $\pm1\%$ can be distinguished with close to perfect accuracy on finite timescales. Based on estimates of nucleation rates for protein condensates in the experiments of Refs.~\cite{Brangwynne2009, Du2018}, we argue that this timescale can be on the order of a few minutes.} 

\section{Results}

\subsection{Nucleation and growth dynamics for ternary fluid}

\begin{figure*}
    \centering
    \includegraphics[width=\linewidth]{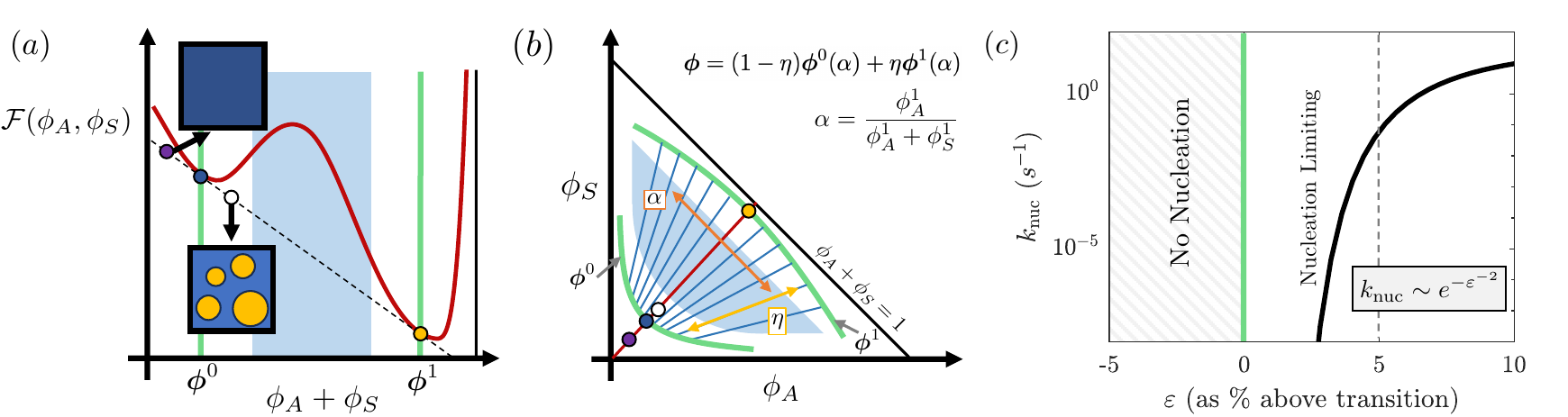}
    \caption{\textbf{Phase diagram and droplet nucleation in a ternary mixture.} (a) Schematic of a double-well free energy from which one can extract the phase equilibria (dark blue is dilute phase, yellow is dense) through the classical common-tangent construction (dashed line). Systems between the green lines  will exhibit phase separation, either through spinodal decomposition (light blue region) or nucleation and growth of droplets (between light blue and green line, e.g.\,the white point). The double-well picture can be understood as a cross-section of the full phase diagram in (b), presented as a function of the concentrations of the two fluids $\phi_A$ and $\phi_S$. We assume we are in a solvent dominated regime and $\phi_A$ and $\phi_S$ are small. We can map the dynamics of the full ternary mixture to that of $(a)$ through defining the dense composition $\alpha$ and dense phase fraction $\eta$, both of which are determined from the point in phase space. An isolated ternary mixture evolves along the tie-line (blue lines) fixed by the initial value for $\alpha$. (a) shows the case of $\alpha=1/2$ denoted by a red tie-line in the (b). \add{(c) Nucleation rate defined in \eqref{eq:nucleation} plotted as a function of supersaturation $\varepsilon=\phi_A-\phi^0_A=\phi_S-\phi^0_S$ expressed as a percentage of $\phi_A^0=\phi_S^0$ (e.g.\,in case of equal $A$ and $S$ concentrations, $\alpha=1/2$).
      When concentrations $\bphi$ are close to the dilute phase equilibria $\bphi^0$, $\varepsilon$ is close to zero and nucleation of droplets becomes very rare.} }
    \label{fig:phaseeq}
\end{figure*}

\add{We consider an incompressible ternary mixture of fluids $A$, $S$ and a solvent in the canonical ensemble with volume fractions} $\phi_A$, $\phi_S$ and $\phi_H=1-\phi_A-\phi_S$, respectively. We consider the scenario where $A$ can condense and form a dense phase only in conjunction with $S$. Here we consider the simplest model for the formation of biomolecular condensates consisting of $A$ and $S$. In response to heat stress, for example, $A$ represents RNA while $S$ models poly(A)-binding proteins in the formation of stress granules (Pab1 in yeast, PABPC1 in humans) \cite{Riback2017, Yoo2019}.

We determine the coexisting phase densities through the Flory-Huggins theory of mixtures, in which the free energy per unit volume for $\bphi=(\phi_A,\phi_S)$ is given by \cite{Mao2019}: 
\begin{equation}
\mathcal{F}_{\rm FH}(\bphi)=RTc_0\left[\sum_{i=A,S}\phi_i\ln(\phi_i) + \frac{\phi_H}{v_s}\ln(\phi_H) +\chi\phi_A\phi_S\right],
\end{equation}
\add{where $c_0$ is the total concentration of solutes, and $v_s$ is the (dimensionless) volume ratio between solvent and soluble molecules and $\chi<0$ sets the strength of the attractive interaction present between A and S molecules.} We set $v_s=1$ below. The first two terms describe the mixing entropy and the last term describes the interaction energy between the two types of molecules. \add{In what follows, we non-dimensionalize the free energy by rescaling with the characteristic energy density $RT c_0$, where $R$ is the gas constant and $T$ is temperature, thus setting $RT c_0 = 1$ in our equations and establishing a natural thermal energy scale (per unit volume). }

\add{For sufficiently negative $\chi$ (we set $\chi=-12$ for the simulation results discussed below), the system phase separates into co-existing dense and dilute phases, the dense phase being rich in $A$ and $S$, while the dilute phase is rich in solvent: the instability of the homogeneous mixture corresponds to a negative determinant for the Hessian of the free energy. }  
We denote the coexisting phase densities by  $\bphi^0 = (\phi_A^0, \phi_S^0)$ and $\bphi^1 = (\phi_A^1, \phi_S^1)$ for the dilute and dense phases, respectively (Fig. \ref{fig:phaseeq}(a)). To determine $\bphi^0$ and $\bphi^1$, we write the free energy density of the two-phase system (for now ignoring interfacial energies) as $\mathcal{F}_{\rm PS}=\eta\mathcal{F}_{\rm FH}(\bphi^1) + (1-\eta)\mathcal{F}_{\rm FH}(\bphi^0)$ where we have defined $\eta$ as the dense phase volume fraction. Minimising this with respect to $\eta$, $\bphi^0$ and $\bphi^1$ gives us the phase equilibria. The resulting equations defining this optimisation problem are classically interpreted in the following way: the chemical potentials of each component, defined through $\mu_i = {\delta \mathcal{F}_{\rm FH}}/{\delta \phi_i}$ (where $i=A, S$), and the pressure, defined as $P = \mathcal{F}_{\rm FH}(\bphi) - \bphi \cdot \bm{\mu}$ \add{(where we recall the scalar product $\bphi \cdot \bm{\mu}=\sum_i \phi_i \mu_i$)}, must equate between the two phases.

Equating $\mu_A,\: \mu_S$ and $P$ provides three equations, but there are 4 unknown quantities in $\bphi^0$ and $\bphi^1$. Crucially, the convex hull of the free energy surface is in fact a plane that must be parametrized by a family of chords (Fig. \ref{fig:phaseeq}(b)):
this parametrization is done through the relative fractions of $A$ and $S$ in the dense phase: 

\begin{equation}\label{eq:alpha}
    \alpha = \frac{\phi_A^1}{\phi_A^1+\phi_S^1}.
\end{equation}
This will give the desired families of pairs of points, $\bphi^0(\alpha)$ and $\bphi^1(\alpha)$. Finally, we identify another equation satisfied by the phase equilibria and $\eta$ due to the conservation of mass:
\beq\label{eq:dpfraction}
\bphi=(1-\eta) \bphi^0(\alpha)+\eta \bphi^1(\alpha).
\eeq
where $\bphi$ is the supersaturated concentrations of $A$ and $S$ (i.e.  before a droplet forms). 
In total, this gives us now 6 equations which can be solved to determine the 6 unknown quantities: $\alpha$, $\eta$ and the four coexisting densities.  In the limit of strong interactions, the dense phase becomes very dense and the dilute phase very dilute. In this limit it is possible to derive analytic forms for the coexisting densities (see details in \seeSI).

Fig. \ref{fig:phaseeq}(a) illustrates a schematic phase diagram for a classical binary mixture and represents a cross section of the full picture for our ternary mixture given in Fig. \ref{fig:phaseeq}(b): in each case we highlight the coexisting densities in green. We denote the phase equilibria by dark blue (dilute, $\bphi^0$) and yellow (dense, $\bphi^1$) dots. For a well-mixed system initialised at concentrations $\bphi$ in the region outside of the green lines (e.g.\,at the purple dot) no phase separation will occur.
 The light blue region denotes when the homogeneous mixture is unstable to fluctuations: here, phase separation occurs spontaneously through so-called spinodal decomposition. Between the light blue region and the green lines (e.g. white point), droplet formation requires the system to overcome an energy barrier due to surface tension when nucleating droplets (that barrier disappears in the light blue region). Following nucleation droplets grow deterministically.
\add{This phase separation mechanism is referred to as nucleation-and-growth.} 

We want to use the sharp transition at $\bphi^0$ 
to distinguish concentrations. Near this boundary, nucleation-and-growth is the only mechanism for phase separation, thus it constitutes the primary focus of our model below. In the current work, we do not consider the case where two systems are at very different concentrations (e.g.\,one system is exhibiting spinodal decomposition and the other nucleation-and-growth) as these are much simpler cases to distinguish (the time- and length-scales associated with phase separation are very different). More broadly, our choice to focus solely on nucleation-and-growth dynamics may actually apply to a wide range of natural systems: it has also been argued recently that the nucleation-and-growth regime in parameter space expands with the complexity of a fluid mixture \cite{Qiang2025} thus making it the dominant mechanism driving phase separation in multicomponent systems \cite{Jacobs2013, Riback2020, Jacobs2021}.

The rate at which droplets nucleate is set by the height of the nucleation energy barrier. 
From the free energy density $\mathcal{F}_{\rm FH}$, supplemented by a surface energy term, we can calculate the energetic difference between a system with and without a droplet of radius $R$ at composition $\alpha$,
$\Delta F(R)=v(P(\bphi^1(\alpha))-P(\bphi))+\gamma s$, where $v=(4/3)\pi R^3$ is the droplet volume, $s=4\pi R^2$ its surface area, $\gamma$ the surface tension, and $P(\bphi)$ the pressure (\seeSI). Crucially, our assumption is that the critical droplet has the same composition $\alpha$ of the phase equilibria: in principle, the saddle-point in the energy landscape describing the minimal energy barrier between the two states may appear at a different composition, but we do not consider this here.
The energy difference $\Delta F(R)$
is non-monotonic in $R$ and is maximised at a critical radius of $R_c$.
Droplets that form with a radius smaller than $R_c$ will dissolve: the system will relax back to a homogeneous state. Larger droplets will survive, so we only consider the rate at which droplets of radius $R=R_c$ form as these are the only ones that will persist in the system beyond short times. 

We find two equivalent expressions for the critical radius that maximises $\Delta {F}(R)$ (derived in full in \seeSI). Defining $\Delta \phi_A=\phi_A^1 - \phi_A^0$ and $\varepsilon_A(t) = \phi_A(t) - \phi_A^0$ (and similarly for $S$), we derive
\begin{equation}\label{eq:Rc}
    R_c(\alpha) = \frac{\ell_0 ( \phi_A^0 + \phi_S^0)\Delta \phi_A}{\varepsilon_A} = \frac{\ell_0 ( \phi_A^0 + \phi_S^0)\Delta \phi_S}{\varepsilon_S},
\end{equation}
\add{where $\ell_0$ is the capillary length (see mathematical definition in \seeSI), the characteristic lengthscale at which curvature‐dependent shifts in chemical potential, as given by the Gibbs-Thomson relation, become significant.} \add{Note $R_c$ is implicitly time-dependent when concentrations $\phi_A$ or $\phi_S$ are depleted upon forming droplets (which enforces that $\varepsilon_A$ and $\varepsilon_S$ decrease in time).} Under the assumption that the interactions between $A$ and $S$ molecules are very strong, one can assume that the dense phase is very dense (and the dilute phase very dilute) in which case this capillary length is simply $\ell_0 =  2\gamma$.
At first, it appears that the last equality in \eqref{eq:Rc} is not necessarily satisfied, but it can be seen through \eqref{eq:dpfraction}, which enforces that the dense volume fraction $\eta$ satisfies $\eta = \Delta \phi_A/\varepsilon_A=\Delta \phi_S/\varepsilon_S$. 
The maximum of the energy barrier reads:
\begin{equation}\label{eq:delF}
    \Delta {F}(R_c) = {v(R_c)}\frac{2 \gamma}{R_c} + \gamma s(R_c) = \frac{4 \pi \gamma R_c^2}{3}. 
\end{equation}
Following the standard approach from classical nucleation theory \cite{Oxtoby1992}, we use this energy barrier to define an approximate nucleation rate via Arrhenius' law in the form
\begin{equation}\label{eq:nucleation}
    k_{\rm nuc} = k_0 V\exp\left[-\Delta {F}(R_c)\right] = k_0 V\exp\left[-4\pi\gamma R_c^2 / 3\right], 
\end{equation}
where $k_0$ is assumed to be a constant pre-factor, independent of the compositions or concentrations in the system \add{and $V$ is the volume of the system.} The idea here is that the exponential term dominates the rate, so ignoring microscopic details in the pre-factor has a negligible effect on the precise value of the nucleation rate.

\add{We omit in our nucleation model here the transient phase of fast initial droplet growth predicted theoretically by Wagner \cite{Wagner1961}: the assumption amounts to modelling that the nucleation is dominated precisely by the crossing of the energy barrier modelled in our rate Eq. [6]. In binary fluids at small supersaturation $\varepsilon$, this nucleation timescale scales like $\exp(\varepsilon^{-2})$, whereas the transient timescale for the Wagner regime is set by diffusion through the droplet, thus scales like $1/\varepsilon^3$, thus is negligible in comparison.}

Once a droplet (indexed by $j$) nucleates, diffusive fluxes drive material into the droplet. We thus require a dynamical description for the growth of a droplet of size $R_j$ and composition $\alpha_j$. Following a now standard approach (see full details in \seeSI and e.g.\,Refs.\,\cite{Zwicker2015, Weber2019}), we arrive at our growth equation for a spherical droplet of radius $R_j$ and composition $\alpha_j$:
\begin{equation}\label{eq:dRdt}
\frac{dR_j}{dt} = \frac{D \ell_0 (\phi_A^0 + \phi_S^0)}{R_j}\left[\frac{1}{R_c(\alpha_j)} - \frac{1}{R_j}\right],
\end{equation}
where $D$ is the diffusion coefficient for an isolated molecule in the dilute phase and $R_c(\alpha_j)$ is given by \eqref{eq:Rc}. 
Note that $\alpha_j$ defines the composition of a specific droplet: it may differ between droplets and from the composition $\alpha$ defined above for the supersaturated phase. Similarly, the critical radius here varies between droplets due to the dependence on the droplet composition $\alpha_j$. Droplets with composition $\alpha_j$ need to have a radius greater than $R_c(\alpha_j)$ to grow in the system; smaller droplets will shrink and dissolve. While we do not model direct interactions (e.g.\,coalescence) between droplets, the critical droplet size for all droplets will increase as the growth of droplets leads to the depletion of $\phi_A(t)$ and $\phi_S(t)$. This leads to antagonistic effective interactions between droplets, modelling the effect of Ostwald ripening. 

The change in the composition of each droplet can be derived in a similar manner to \eqref{eq:dRdt} and takes the form (\seeSI)
\begin{equation}
    \frac{d\alpha_j}{dt} = \frac{3D}{R_j^2} \left[\frac{\Delta\phi_S\varepsilon_A - \Delta\phi_A\varepsilon_S}{\Delta\phi_A + \Delta \phi_S} \right]\label{eq:dadt}
\end{equation}
where crucially the $\bphi^0$ and $\bphi^1$ that appear here in the terms \,$\Delta \phi_A =\phi^1_A - \phi_A^0$ and $\varepsilon_A =\phi_A - \phi_A^0$ are evaluated from the droplet's composition $\alpha_j$, not that of the phase equilibria for $\phi_A$ and $\phi_S$ (namely $\alpha$), so the right-hand side can be non-zero when $\alpha \ne \alpha_j.$

We now have a complete description of droplet nucleation and growth in a ternary mixture. The dynamics for droplet nucleation and growth at time $t$ are set by the (non-equilibrated) concentrations $\bphi(t)$ outside the droplets. We study the dynamics of the ternary mixture through numerical simulations of our dynamical model, the details of which are given in \seeSI. 

To compare the results of the simulations to typical biomolecular condensates, we set the parameters of our model:
we fix the system volume $V=1000\,\mu m^3$, comparable to that of a one cell embryo of 
{\em C. elegans}
\cite{Brangwynne2009}, and $D=1\,\mu m^2 / s$, a typical diffusion coefficient for small molecules in the cell \cite{Mine2021}. We are then left to set the surface tension $\gamma$ and the pre-factor to the nucleation rate $k_0$. Experimental work measuring the surface tension of biomolecular condensates estimate it between $10^{-4} -10^{-7}$ of that of the surface tension between air and water \cite{Brangwynne2009}. We set $\gamma = 10^3 k_BT / \mu m^2\approx 10^{-4}\gamma_{\rm air-water}$. Finally, we look to set $k_0$. Microscopic formulations of the nucleation rate prefactor have received much attention \cite{Kramers1940, Oxtoby1992}, but an exact treatment remains beyond the scope of this work. Instead, we set $k_0$ by taking as a typical system one where there is a supersaturation of $10\%$ for both $A$ and $S$. We choose $k_0$ such that this level of supersaturation drives the formation of $\sim10-100$ droplets on a timescale of minutes, with the coarsening to a single droplet on the timescale of hours \cite{Du2018}. The number of droplets is set by the ratio of $D$ and $k_0$: if growth is much faster than nucleation, we can expect only a few droplets to form. Conversely, slow growth allows for many droplets. In this way, we choose $k_0V=50s^{-1}$ such that $k_0 = 0.05 s^{-1} \mu m^{-3}$. Finally, {we set the maximal decision time to $T_f=10$ minutes. Different decision making times $T_f$ are explored in the \seeSI (Fig. S2): we observe quantitatively comparable results for other $T_f$ on the order of minutes.}

\subsection{Rare nucleation hinders sensing near phase transition}

\begin{figure}
    \centering
    \includegraphics[width=\linewidth]{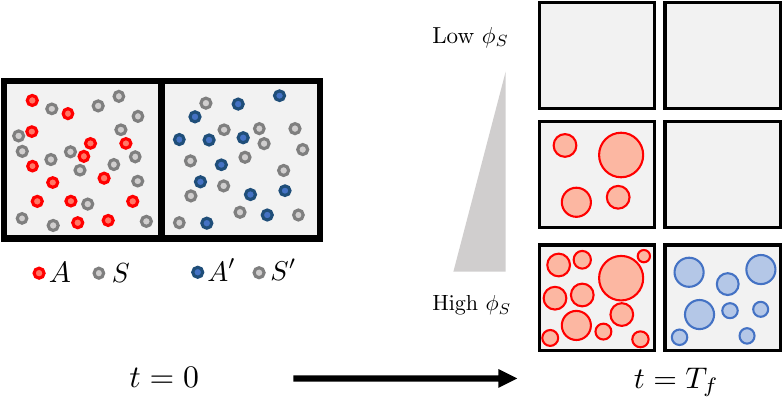}
    \caption{\textbf{Using a second system as a measuring stick.} 
    Schematic for dynamics of two ternary mixtures, identical except for initial concentrations of $A$ (red) and $A'$ (blue). The amount of $S$ determines the relative volumes of condensates. Too little $S$ prevents formation of droplets before the decision time $T_f$, whereas too much $S$ renders initial differences between $A$ and $A'$ insignificant.}
    \label{fig:second}
\end{figure}

Suppose that a cell wants to determine whether the concentrations of $A$ and $S$ molecules are above or below a threshold set by the transition point to a phase-separated state (dark blue point in Fig. \ref{fig:phaseeq}(a),(b)). 
We say that a cell decides they are above or below if a droplet nucleates or not before a fixed decision time. Here and below, we will refer to this as \textit{first-nucleation} sensing. 

From \eqref{eq:Rc}, we see that the critical droplet radius is large for concentrations $\bphi$ close to $\bphi^0$. This results in a very low nucleation rate through \eqref{eq:nucleation}. Physically, this is due to increased energy barriers to escape what is initially a metastable state. These barriers pose a challenge to systems using droplet formation to sense concentrations near the transition point. 

To illustrate this mathematically, we consider the singular case of equal initial concentrations for $A$ and $S$, in which the equations simplify: \add{in this case, the phase equilibria are the same and $\phi_A^0=\phi_S^0\approx 0.025$.} We define a single variable $\varepsilon = \varepsilon_{A/S}=\phi_{A/S}-\phi_{A/S}^0$ capturing the initial supersaturations of both $A$ and $S$. From \eqref{eq:Rc} and \eqref{eq:nucleation}, the timescale for the first nucleation event scales like $\tau_1=k_{\rm nuc}^{-1}\sim\exp\left(\varepsilon^{-2}\right)$, as is demonstrated in Fig. \ref{fig:phaseeq}(c). This timescale diverges faster than exponentially in the limit of small supersaturation $\varepsilon \rightarrow 0$. (A similar scaling relation can be argued for the general case of $\phi_A\ne \phi_S$.) 

\add{Phase separation alone cannot accurately discriminate concentrations close to the transition point in finite time due to rare nucleation events. We can quantify this for our ternary mixture: for a decision time of $T_f=10$ minutes and equal $A$ and $S$ concentrations, we see from Fig. \ref{fig:phaseeq}(c) that we would require a supersaturation of at least  $\sim5\%$ in $A$ and $S$ for a nucleation rate $k_{\rm nuc}$ such that $k_{\rm nuc}T_f\gg 1$.  }

\add{Another illustrative case is when $\phi_{A}=\phi_{A}^0=\phi_{S}=\phi_{S}^0$ initially and we ask how much $\phi_A$ would need to increase to escape the nucleation-limiting regime. We deduce that it is a larger increase than $5\%$ of $\phi_A$: from Fig. \ref{fig:phaseeq}(b), we see that increasing $\phi_A$ alone would lead to a decrease in $\phi_S^0$ and an increase in $\phi_A^0$. Taking these changes to be linear, we argue that $\phi_A$ would need to increase by $\approx10\%$ to realize $5\%$ supersaturations in $A$ and $S$ and escape the nucleation-limiting regime for $T_f=10$ minutes.} 

For smaller supersaturation, this sensing mechanism suffers from false negative results: no nucleation before a finite decision time despite being in the phase-separating regime. This suggests upper bounds on the precision with which phase separation alone can discriminate concentrations. The bound is comparable to that derived for hunchback promoters performing a read-out of the bicoid morphogen gradient, where cell sense concentration differences in signalling molecules on the order of 10$\%$ in minutes \cite{Gregor2007, Desponds2020}, despite the two sensing mechanisms being very different.

One route to circumvent this would be to fix a decision time $T_f$ and then find the supersaturation $\varepsilon(T_f)$ at which we could expect to see a droplet form (i.e.\,with probability $\approx1/2$). We could then use phase separation to signal whether $\phi_A > \phi_A^0+\varepsilon_A(T_f)$ (and similarly for $\phi_S$) effectively shifting the transition. However, there are several difficulties with implementing this approach: this new effective critical concentration changes with decision time $T_f$, so accurate decisions now require a strict measure of time for the sensing process, an added layer of computation. Also, the shifted critical concentration can generate false positives: seeing phase separation when $\phi_A^0<\phi_A<\phi_A^0 + \varepsilon_A(T_f)$.

\subsection{Using a second subsystem as a measuring stick: overcoming rare nucleation}

We have demonstrated that sensing concentrations near the transition point can be inaccurate due to rare nucleation. Now, we move beyond using the formation of droplets \textit{alone} as a sensor for inferring concentrations: we consider a second subsystem, which itself is also a ternary mixture. We assume the second subsystem is identical to the original one, apart from the initial concentration of $A$, which we label as $A'$ for the second subsystem. \add{For now, we will assume that the two subsystems evolve independently, then consider interactions between subsystems in Sections D and E below.} For consistency, we denote the scaffolds by $S$ and $S'$ across the two systems (see Fig. \ref{fig:second}). \add{We will keep the initial concentration of $A'$ constant at $\phi_{A'}=0.025$, while varying that of $A$ in the first subsystem. This puts us in the parameter regime where the composition of droplets is roughly equal in $A$ and $S$.}

\add{The idea is to now measure the concentration of $A$ \textit{relative to} $A'$, such that  $A'$ represents a measuring stick for $A$. (In reality, the picture could be far more sophisticated, say $A'$ triggers a process with the opposite functionality to the one triggered by $A$ and thus the two concentrations serve as measuring sticks for each other.) Measuring sticks can be implemented through simple binding, where a sensor molecule binds to its target and changes its properties, or through competition, where two molecules compete for the same binding site and the balance shifts with concentration. Such mechanisms resemble chemical titrations, where binding equilibria determine the proportion of sensor molecules in each state. Here, we apply this idea in the context of droplets: we propose that the volume of droplets (which play the role of the measurable signal here) can be used to infer relative concentrations.}

We also update our decision-making mechanism: a cell samples a molecule at random from the droplets and determines whether $A$ or $A'$ is more abundant from whether it samples an $A$ or $A'$ molecule. The probability of this is simply the ratio of the droplet volumes weighted by the probability that a nucleation event occurs. We refer to this mechanism as \textit{droplet-proportion} sensing and argue that it represents the simplest mechanism to investigate how phase separation may amplify concentration differences.  More sophisticated sensing might incorporate spatial and temporal information, for example, leading to more accurate sensing, but this is beyond the scope of the current work.

The usefulness of a second phase-separating system can be illustrated through the following argument: one immediate consequence of comparing two droplet-forming systems is that concentration differences are amplified in the dense phases. Randomly sampling an $A$ or $A'$ molecule from the whole system when both subsystems are well-mixed would mean a probability of $\phi_A / (\phi_A +\phi_{A'})$ of picking an $A$ molecule (vs $A'$). As a sensing mechanism, this probability describes a Hill curve (or Monod equation) with a maximum slope equal to $1/(4\phi_{A'})$ when $\phi_A=\phi_{A'}$ which defines an effective measure of sensitivity. Droplet-proportion sensing is much more precise. {Through a lever rule argument, we expect} a system initialised with concentrations $\bphi$ to nucleate droplets, growing until the concentration in the dilute phase is close to $\bphi^0$. Ignoring interfacial effects, the total volume of the dense phase would be approximately set by the supersaturations as $(\varepsilon_A+\varepsilon_S)V\approx 2 {\varepsilon_A} V$ for comparable concentrations of $A$ and $S$.
\add{Droplet-proportion sensing would then imply a probability ${\varepsilon_A}/({\varepsilon_A + \varepsilon_{A'})}$ of picking $A$
corresponding to an effective sensitivity of $1/4\varepsilon_{A'}$, much greater than that of sampling the initial mixture (provided $\varepsilon_{A'}<\phi_{A'}$ which must hold because $\phi_{A'}^0>0$). Clearly this (equilibrium) sensitivity can be made large through reducing the supersaturation, as opposed to reducing the overall concentration of $A$ in the absence of droplets, enabling accurate sensing at moderate concentrations. 
 }
While this lever rule argument predicts the sensitivity expected at equilibrium, we demonstrate that one can achieve higher sensitivity in nucleation-limited regimes on relevant sensing timescales.

Fig. \ref{fig:competition}(b) displays the results of our numerical simulations for two independent ternary mixtures, where the results are strongly controlled by the concentration $\phi_S$ (parametrising the different curves). More specifically, we recall that $\phi_{S'}^0$ is the concentration of $S'$ required for the $A'$-subsystem to nucleate droplets given the concentration $\phi_{A'}$. In Fig. \ref{fig:competition}(b), we vary the initial concentrations of $S$ and $S'$ through $\varepsilon_{S'}$ as $\phi_S(t=0)=\phi_{S'}(t=0)=\phi_{S'}^0 + \varepsilon_{S'}$ (where the choice of $S'$ label in $\varepsilon_{S'}$ denotes the distance to $\phi_{S'}^0$). We expect that at small $\varepsilon_{S'}$, neither system nucleates before the finite decision time due to the rare nucleation discussed in the previous section, whereas at large $\varepsilon_{S'}$, both system nucleate many droplets (see again the schematic in Fig. \ref{fig:second}).

This is confirmed in the simulations (Fig. \ref{fig:competition}(b)): at low $\varepsilon_{S'}$, the $\bphi'$ system never nucleates because it is too close to the transition line, and only when $\phi_A$ is substantially larger than $\phi_{A'}$ does the $\bphi$ system nucleate: see light blue curve Fig. \ref{fig:second}(b), when $\varepsilon_{S'}= 5\%$ of $\phi_{S'}^0$. As $\varepsilon_{S'}$ increases, the response curve in Fig. \ref{fig:competition}(b) shifts to the left (blue curve, $7\%$). This is due to $\bphi$ systems with $\phi_A>\phi_{A'}$ being able to nucleate droplets as $S$ becomes more abundant. For larger $\varepsilon_{S'}$, the $\bphi'$ system also nucleates droplets, so droplet-proportion sensing becomes less precise --- see the flattening of the response function  as the supersaturation in $S$ above $\phi_{S'}^0$ further increases $20\%$ (dark blue curve in panel Fig. \ref{fig:competition}(b)). {While at small $\varepsilon_{S'}$, the dynamics are nucleation-limited, systems with large $\varepsilon_{S'}$ deplete resources quickly by nucleating and growing droplets. This is illustrated in Fig. \ref{fig:competition}(c), which we will be further discussed below. The large $\varepsilon_{S'}$ curves in Fig. \ref{fig:competition}(b), such as the dark blue curve, provide results comparable to the lever rule argument given above. Our full dynamical model allows us to explore beyond this limit and to compare to finite decision timescales.}

These results point to an optimal $\varepsilon_{S'}$ for this sensing set-up. To quantify the accuracy of droplet-proportion sensing in this scenario, we need to devise a metric. The perfect scenario for this sensing mechanism is one where only droplets of $A$ are present if $\phi_A >\phi_{A'}$ (and only droplets of $A'$ if $\phi_A <\phi_{A'}$). We define an accuracy score by taking the difference between a step function centered at $\phi_A=\phi_{A'}$ and the $A$-fraction droplet volumes in Fig. \ref{fig:competition}(b), and integrating this error score with a Gaussian function centered at $\phi_A=\phi_{A'}$ with standard deviation of $2\%$ of $\phi_{A'}$. The total accuracy is then 1 minus the integral.  
This accuracy quantifies with what probability does phase separation and droplet-proportion sensing correctly conclude whether $\phi_A$ is larger or smaller than $\phi_{A'}$ when the two concentrations are very similar to each other. An accuracy of $1$ indicates perfect ability to sense concentrations relative to $\phi_{A'}$, whereas an accuracy of $0.5$ implies the sensing mechanism is as useful as flipping a coin. The accuracy is plotted in Fig. \ref{fig:competition}(d): we observe a peak accuracy of around $80\%$ for a supersaturation $\varepsilon_{S'}\approx 7$-$8\%$. 

The optimal value for $\varepsilon_{S'}$ can be reasoned as follows. As for a single ternary mixture, the finite decision time $T_f$ imposes an effective (non-zero) boundary for the supersaturation below which we expect to not see droplets (recall the \textit{nucleation-limiting} regime in Fig. \ref{fig:phaseeq}(c)). The optimal $\varepsilon_{S'}$ sits just below this boundary for the $\bphi'$ system, such that whenever $\phi_A>\phi_{A'}$, we would expect the $\bphi$ subsystem to nucleate droplets, but not the $\bphi'$ subsystem. For smaller $\varepsilon_{S'}$, neither subsystem reliably nucleates droplets, whereas large $\varepsilon_{S'}$ ensures both nucleate {and the ratio of droplet volumes is comparable to the lever rule argument given above.} For our decision time $T_f=10$ minutes, we argued above for a supersaturation $\varepsilon_{S'}\geq 10\%$ of $\phi_{S'}^0$ for the $\bphi'$ subsystem to nucleate. This agrees with our observation of an optimal supersaturation in Fig. \ref{fig:competition}(d) of around $7$-$8\%$ for the orange points. 

\subsection{Interactions between subsystems: Competition for building blocks $S$ strengthens sensitivity}

\begin{figure*}[ht!]
\centering
\includegraphics[width=\linewidth]{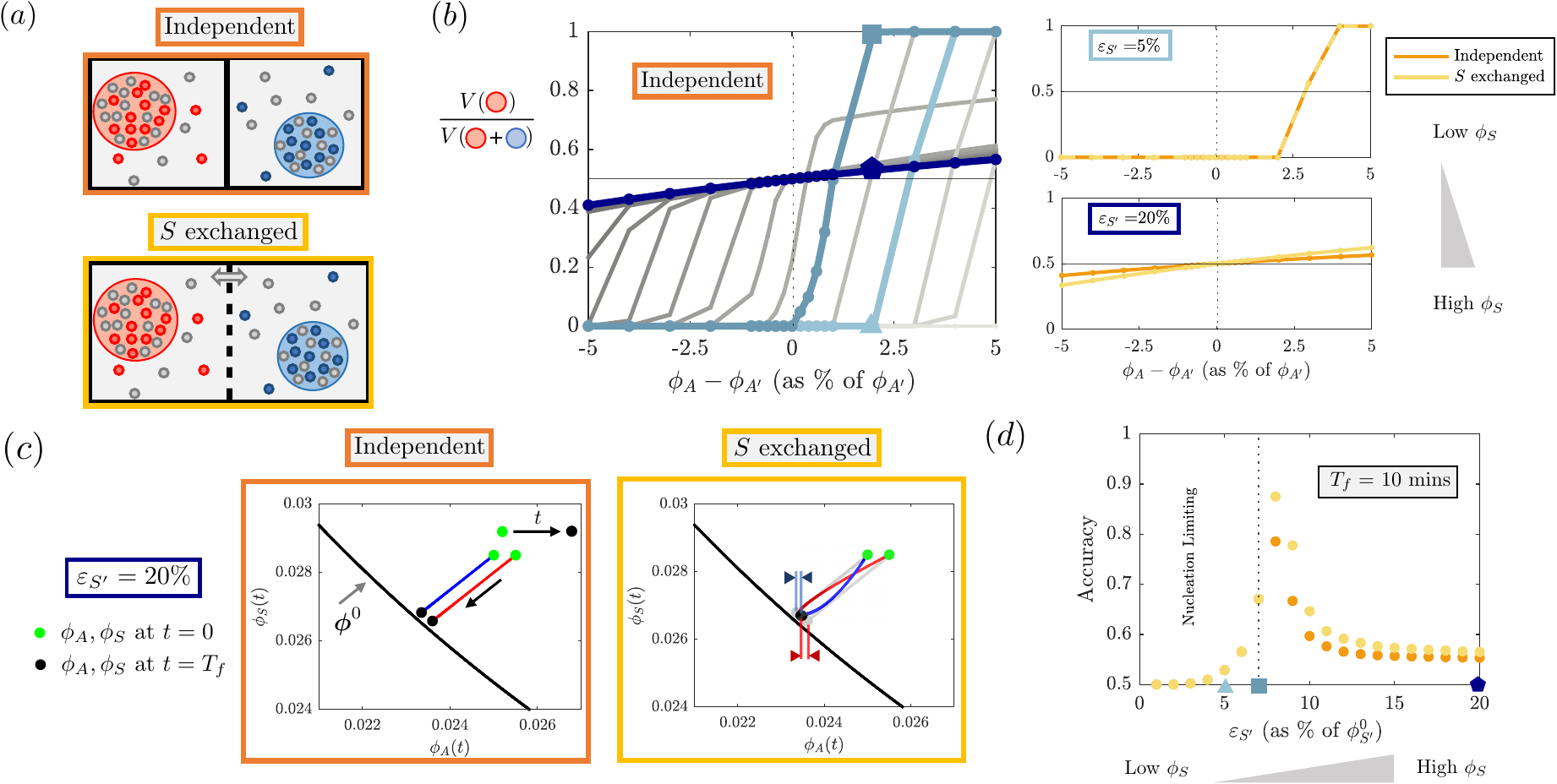}
\caption{\textbf{Competition for $S$ increases sensitivity.} (a) We compare the dynamics of two ternary mixtures without (orange) and with (yellow) exchange of $S$. \add{(b) For the two independent mixtures, the proportion of red droplets at time $T_f$ as a function of the initial difference in $A$ and $A'$ concentrations is plotted. Each curve is a different supersaturation of $S$, denoted $\varepsilon_{S'}$ and measured as a percentage of $\phi_{S'}^0$: the curves shift to the left and flatten (light blue $\rightarrow$ blue $\rightarrow$ dark blue) as the initial supersaturation of $S$ increases from $0\%$ to $25\%$ of $\phi_S^0$ (colored lines from $5\%\rightarrow 7\%\rightarrow 20\%$, respectively). A comparison is made in the rightmost panels to the case with exchange of $S$ between subsystems (in yellow): the two scenarios are similar at small $\varepsilon_{S'}$ (light blue) but different at large $\varepsilon_{S'}$ (dark blue).} \add{(c) We explain this through phase space trajectories for the (non-equilibrated) dilute phase concentrations, $\bphi(t)$ in red and $\bphi'(t)$ in blue. For the independent systems, these evolve along their respective tie-lines. In the case of $S$ exchange, initial growth is faster in the $\bphi$ subsystem, driving a flux of $S$ from the $\bphi'$ to the $\bphi$ system due to depletion. Eventually, the $\bphi'$ system grows droplets until there is no flux of $S$ between the systems and they share the same dilute phase concentrations. The difference between the final state is highlighted between the two scenarios. The uptake of $A$ to the dense phase is proportional to the difference in the $x$-direction between the start (in green) and end (black) points. Competition enforces larger uptake of $A$ in the red system, thus a larger dense phase, explaining the steeper curve in (b).} (d) When employing \textit{droplet-proportion} sensing, an optimal supersaturation of $\varepsilon_{S'}\approx 7$-$9\%$ is found to maximise the accuracy at $80\%$ for the independent systems and $ 90\%$ when exchanging $S$. \add{The line at $\varepsilon_{S'}=7\%$ denotes the point above which the exchange of $S$ makes a difference.} 
 }
\label{fig:competition}
\end{figure*}

So far the two subsystems have been assumed to evolve independently of one another. We now let $S$ be exchanged between the two subsystems (see schematic in Fig.\,\ref{fig:competition}(a)). \add{We assume that any difference in the instantaneous concentration of $S$ in the supersaturated dilute phase $\phi_S(t)$ quickly relaxes by exchanging $S$, ensuring $\phi_S(t)$ is equal across the two systems at all times. We choose not to fix a finite timescale for this exchange of S: this assumes that the two subsystems draw upon the same pool of S molecules. Our model of two ternary mixtures thus approximates a \textit{quaternary} mixture ($A$, $A’$, $S$ and solvent) where $A$ and $A’$ have a strong negative interaction. While a thermodynamic description for this mixture would be interesting, there is added complication due to the coexistence of $A$-rich and $A’$-rich dense phases. We thus employ the ternary mixtures (augmented with fast exchanges of $S$) as a proxy for this more complex mixture.}

Numerical simulations conclude that competition for $S$ results in a stronger sensitivity compared to the independent systems at larger supersaturation of $S$, as illustrated by the increased slope of the sensitivity curve in Fig.\,\ref{fig:competition}(b) at $\phi_A=\phi_A'$ for $\varepsilon_{S'}=20\%$ of $\phi_{S'}^0$. The two perform equally well at small supersaturation as they are both limited by rare nucleation (see Fig.\,\ref{fig:competition}(b) panel with $\varepsilon_{S'}=5\%$). This enhanced sensitivity is explained graphically through the phase portraits in Fig.~\ref{fig:competition}(c). In the absence of $S$ exchange, the dilute concentrations converge towards the equilibrium dilute phase concentrations (albeit not exactly due to system finite size effects) {leading to a ratio of droplet volumes comparable to the one predicted by a lever rule argument}. The slope of these trajectories is set by $\alpha$ (which is constant when the two systems evolve independently).

Exchanging $S$ between subsystems results in the two trajectories converging to the same point. If both subsystems nucleate droplets, the dilute concentrations will converge towards the black curve Fig.~\ref{fig:competition}(c) of dilute phase equilibria. However, the subsystem with a larger excess of $A$ and $S$ will nucleate droplets more quickly, thus demanding excess $S$ from the other subsystem. The subsystem with larger supersaturation will deplete its resources for forming droplets more quickly, at which point the slower subsystem will demand any excess $S$, until the dilute concentrations of $A$ and $S$ are equal across the two systems. Both systems stop forming droplets when their trajectories reach the black transition curve. Since their $S$ concentration is the same, so must be $\phi_A$ and $\phi_{A'}$ (black dot).
It follows, upon comparison to the case without competition, that the more supersaturated system (red curve in Fig. \ref{fig:competition}(c)) is able to form a larger volume of droplets. We see this from the increased reduction of $A$ in the dilute phase. The blue system loses out, and competition for $S$ produces a steeper slope in Fig. \ref{fig:competition}(b) at large $\varepsilon_{S'}$, strengthening sensitivity. {We expect that a revised application of the lever rule, now accounting for the exchange of $S$, could predict the ratio of droplet volumes in this limit of large $\varepsilon_{S'}$}. 

The convergence of $\bphi$ and $\bphi'$ at long times has another physical significance: given that the phase equilibria in a subsystem are determined by the supersaturated concentrations $\bphi$ alone, this implies equality in the phase equilibria across subsystems. It follows that this convergence in $\bphi$ and $\bphi'$ permits the sustaining of dense droplets at steady-state across the two subsystems. Coarsening through classical Ostwald ripening will eventually result in 2 droplets, one in each subsystem, at long times. We stress that this is only possible as $A/A'$ is not exchanged between subsystems. If this was the case, the same Ostwald ripening mechanism would enforce only a single droplet {(comprised of $A$, $A'$ and $S$)} survives at steady-state.

\subsection{Replenishing of $A$ optimises sensing} \label{sec:optimal}

\begin{figure*}
    \centering
    \includegraphics[width=\linewidth]{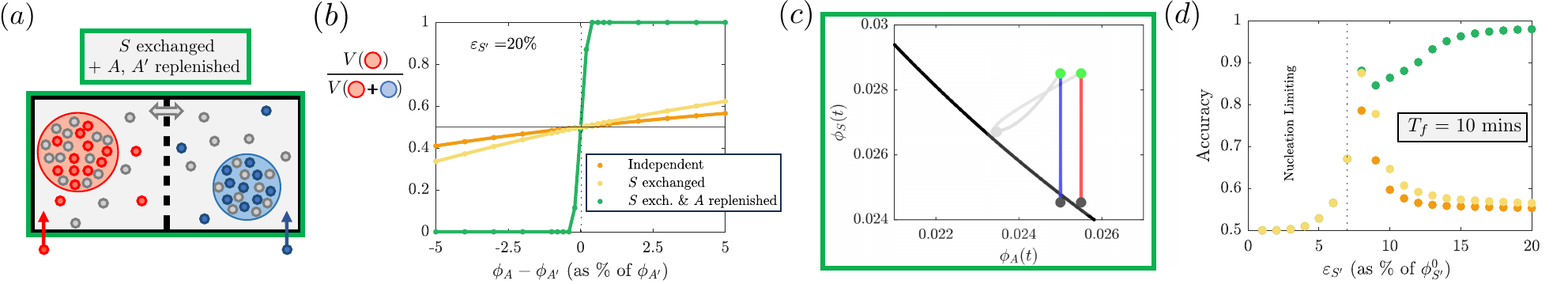}
    \caption{\textbf{Replenishing of $A$ drives optimal sensing.} (a) Schematic for two systems exchanging $S$ and replenishing $A$ and $A'$ (i.e.\,keeping the dilute phase concentration of $A$ and $A'$ constant). (b) Replenishing of $A$ leads to an equilibrium (long-time) state with a homogeneous make-up of condensates. For finite decision-making times, we still see a strong improvement compared to previous scenarios. (c) Replenishing of $A$ confines the phase-space trajectories to vertical lines: nucleation and growth of droplets in the red system will eventually push $\phi_S$ below the phase equilibria for the blue system, ensuring blue droplets are only present transiently. (d) A quantification of how accurately each scenario can sense relative concentrations through forming droplets. When there is enough $S$ for the $A'$ subsystem to nucleate droplets before the decision time, competition for $S$ and replenishing of $A$ is optimal and largely insensitive to supersaturation $\epsilon_S$.}
    \label{fig:optimal}
\end{figure*}

We have seen that a necessary condition for both systems to support droplets when $S$ is exchanged is the convergence of the dilute concentrations $\bphi=\bphi'$. This suggests a simpler mechanism for increasing sensitivity: maintaining the  concentrations $\phi_A$ and $\phi_{A'}$ in the dilute phases (see a schematic in Fig. \ref{fig:optimal}(a)). By doing so, the phase space trajectories will never meet, as they are confined to 2 different vertical lines (Fig. \ref{fig:optimal}(c)). As such, the long-time state of the system does not support the coexistence of droplets. This suggests that replenishing of $A$ and $A'$ upon forming droplets allows for a definitive binary decision through phase separation. 

Numerical simulations confirm our intuition. We observe that phase separation augmented with competition for $S$ and replenishing of $A$ and $A'$ is able to produce very accurate readings of whether $\phi_A>\phi_{A'}$ in Fig. \ref{fig:optimal}(b).
At equilibrium, we necessarily require $\phi_S\rightarrow \phi_S^0(\phi_A)$ (black dot at the end of the red line in Fig. \ref{fig:optimal}(c)). In doing so, we drive $\phi_S(t)$ in the $\bphi'$ subsystem below the dilute phase equilibria, forcing this subsystem out of the phase separating regime (black dot at the end of the blue line). While droplets may form initially in the $\bphi'$ system, they are purely transient as eventually they will dissolve due to this competition for $\phi_S$.

The picture at equilibrium is thus optimal: a homogeneous dense phase consisting entirely $A$ or $A'$ droplets perfectly reflecting the more abundant of the two. Our simulations of nucleation and growth show that this optimal performance is maintained down to low $\varepsilon_S' \sim 8\%$, below which nucleation limitation reduces accuracy. However, as Fig. \ref{fig:optimal}(d) shows, the reduced performance of this model due to nucleation is never worse than the nucleation-enhanced performance of prior models without replenishment.

To highlight another advantage of this model with replenishment, the abundance of $S$ plays only a minor role,
as confirmed in Fig. \ref{fig:optimal}(d). Provided $\varepsilon_{S'}$ is large enough for both subsystems to nucleate droplets (more than $7\%$ of $\phi_{S'}^0$ for the considered decision time), we see little consequence of having an abundance of $S$ in the system. This is in stark contrast to the previous scenarios where $A$ and $A'$ were not replenished. There, a fine control was required on the choice for the initial concentration of $S$ to maximise the accuracy of the process. \add{The scenario considered in this section, where the two phase separating systems compete for a common pool of $S$ and replenish $A$ upon forming droplets, can in principle be employed by cells to distinguish concentration differences of 1\% in minutes.}

\subsection{Winner-takes-it-all: sensing through first nucleation event}

\begin{figure*}
    \centering
        \includegraphics[width=\linewidth]{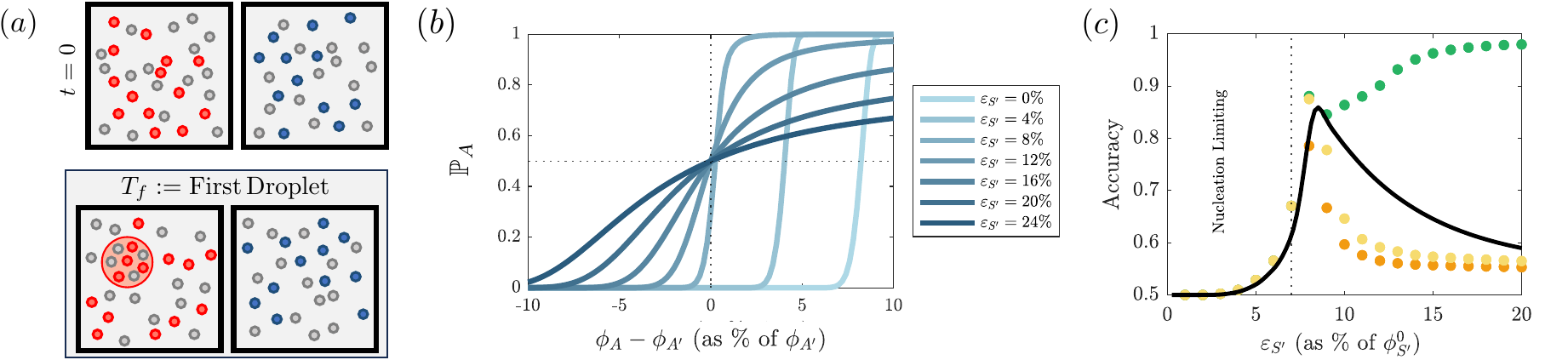}
    \caption{{\textbf{Sensing from first nucleation event.} (a) Schematic of inferring higher concentrations  from the first nucleation event. 
    (b) Probability of the $\bphi$ system nucleating before $\bphi'$ and before decision time $T_f$ as a function of concentration differences between $A$ and $A'$ ($x$-axis) and supersaturation of $S$ for different values of $\varepsilon_{S'}$. (c) Sensing accuracy compared to other scenarios. First nucleation event sensing, plotted in black, is just as accurate as methods including growth of droplets in the nucleation-limited regime (colour scheme same as in Fig. \ref{fig:optimal}(b)). The accuracy peaks at $\approx85\%$ accuracy but requires close control on the supersaturation $\epsilon_S$ to reach this optimum.  }}
    \label{fig:firstnuc}
\end{figure*}

We have explored how the competition for $S$ and replenishing of $A$ and $A'$ allows for accurate sensing of relative concentrations through the nucleation and growth of droplets. We conclude with results from a simpler physical picture. Instead of waiting for droplets to form and resources to deplete, a cell may instead wait only for a single droplet to form. This first nucleation event could trigger downstream processes on timescales much quicker than that of the nucleation and growth process,  effectively making a decision before nucleating another droplet. The first nucleation event is stochastic in nature, so the system will never truly be able to perfectly determine which concentration is larger from the first droplet. Regardless, we  explore how this first-nucleation sensing mechanism performs when comparing multiple systems as schematised in Fig. \ref{fig:firstnuc}(a). 

The probability that one subsystem nucleates before the other and before a fixed, finite decision time $T_f$ is given by a ratio of the nucleation rates in the model. \eqref{eq:nucleation} gives us the nucleation rate of each system, $k_{\rm nuc}(\bphi)$ and $k_{\rm nuc}(\bphi')$, as a function of the critical radius $R_c(\alpha)$,
where the composition $\alpha$ itself depends on the concentrations in each system $\bphi$ and $\bphi'$. The probability $\mathbb{P}_A$ that the subsystem $\bphi$ nucleates before $\bphi'$ before time $T_f$ is then given by 
\begin{equation}\label{eq:result-1nuc}
    \mathbb{P}_A=\frac{k_{\rm nuc}(\bphi)}{k_{\rm nuc}(\bphi)+k_{\rm nuc}(\bphi')} \left[1 - e^{-(k_{\rm nuc}(\bphi)+k_{\rm nuc}(\bphi'))T_f}\right],
\end{equation}
where the term in the square brackets accounts for the fact that no droplets may nucleate before $T_f$.

The accuracy of this first-nucleation sensing mechanism described in \eqref{eq:result-1nuc} is illustrated in Fig. \ref{fig:firstnuc}(b), where we see a similar shifting and flattening of the signal-response-like curves as the supersaturation of $S$ (defined through $\varepsilon_{S'}$) increases. In Fig. \ref{fig:firstnuc}(c), we compare the resulting accuracy with the other sensing mechanisms. As expected, all mechanisms perform equally in the nucleation-limited regime, where the ability of each sensing mechanism is limited by the rare nucleation of $A$ droplets. Surprisingly, when $S$ is more abundant, the first nucleation event does as well, if not better, than droplet-proportion sensing in the absence of $A$ replenishment (orange and yellow curves). This is striking because first-nucleation sensing is happening on a much shorter timescale: in seconds rather than minutes. However, the maximum accuracy is less than that of the optimal method with competition and replenishment described in the previous section, and requires careful tuning of the supersaturation of $S$.

\section{Discussion}

We have outlined  generic constraints and optimality considerations for sensing relative concentration differences through phase separation. Building on classical approaches \cite{Huggins1941, Flory1941, Cahn1958, Weber2019}, we first derived a mathematical model for nucleation and growth dynamics in a ternary mixture of fluids $A$, $S$ and a solvent, where fluids $A$ and $S$ phase separate through attractive interactions with one another. We then demonstrated that phase separation alone has limited accuracy when sensing concentrations near the phase transition due to rare nucleation events within the framework of classical nucleation theory. The energy barrier which the system must overcome to form droplets diverges as the concentrations approach those at the transition. 
Our results indicate a clear accuracy-time trade-off for precise sensing through the formation of droplets in a fluid mixture. From nucleation rate estimates for protein condensates in the experiments of Refs.~\cite{Brangwynne2009, Du2018}, $\varepsilon_A$ and $\varepsilon_S$ that are less than 5-10$\%$ of the transition concentrations will not reliably nucleate droplets within a decision time of 10 minutes. 

We then proposed a solution to overcome this: the introduction of a second subsystem as a measuring stick. In the current work, we assumed that the two subsystems were physically separated, but a more general model might consider a quaternary mixture of $A$, $A'$ and $S$ in a single system. For this case, some repulsive interaction between $A$ and $A'$ would be needed to drive distinct $A$- and $A'$-droplets along the lines of the current work and this extra interaction could  affect the stability of the homogeneous phase. We believe that such a quaternary mixture will resemble the dynamics of the 2 subsystems exchanging $S$ of the current work, but further analysis of how concentration discrimination can occur through phase separation in multicomponent fluid mixtures would be of great interest. 

We first considered the dynamics of two \textit{independent} subsystems, each ternary mixtures described by the same free energy, but at different concentrations of $A$ in the mixture. We asked how accurately a system can sense the difference in concentration in $A$ in each subsystem from sampling a molecule from a random droplet across the two systems. 
We found concentrations could be discriminated with a higher accuracy than at equilibrium if phase separation is in a nearly-nucleation-limited regime. This kinetic regime of high accuracy is defined by an optimal amount of $S$ that results in $80\%$ accuracy for concentration fluctuations of $2\%$.

A further improvement was found by allowing the exchange of $S$ molecules between subsystems (homogenizing the volume fraction of $S$ outside of droplets in each system) which generated a positive feedback mechanism that amplified concentration differences. The exchange of $S$ ensures convergence of the phase equilibria across the two subsystems, ensuring a larger dense phase in the system with more $A$ as demonstrated in Fig. \ref{fig:competition}(c).
These antagonistic interactions between subsystems resemble mRNA competition during p granule segregation \cite{Saha2016b, Lee2013}, stress granule formation \cite{Sanders2020}, and competition for signaling molecules during growth-factor signaling at the plasma membrane \cite{Oh2022-tv}. We highlight how these interactions drive higher sensitivity when distinguishing signals. 

We then proposed a minimal mechanism for optimal sensing of relative concentration through phase separation between two subsystems: competition for $S$ augmented by the replenishing of $A$ upon droplet formation. This replenishing may be realised with a semi-permeable membrane (allowing $A$ and $A'$, but not $S$, to be exchanged with a bath to maintain concentrations) or through different diffusivities in solution ($A$ and $A'$ diffusing faster than $S$). \add{Another potential source of this replenishment comes in the form of active chemical reactions, the role of which are currently under much scrutiny in the formation of biomolecular condensates \cite{Weber2019}. The production of A and A’ molecules through reactions can counteract the depletion due to droplet formation: \textit{active} reactions would in principle allow for kinetics unrestricted by equilibrium thermodynamics. These reactions are also capable of preferentially suppressing or enhancing nucleation rates \cite{Ziethen2023, Cho2023}; these capabilities could be further leveraged to enhance sensing.} At long times, only one subsystem (the one with more $A$) will sustain droplets in this scenario, as these interactions drive one system out of the phase separating regime, as demonstrated in Fig.\,\ref{fig:optimal}(c). 

We confirmed through simulations that this set-up works well also for finite decision times, distinguishing concentration differences of $\pm1\%$ with close to perfect accuracy within 10 minutes for our proposed nucleation kinetics \cite{Brangwynne2009, Du2018}. This timescale is shorter than those reported for other mechanisms: the ultrasensitive response in Cdc25C activation regulated by the phosphorylation of Cdk1, a classic example of Koshland-Goldbeter kinetics \cite{Goldbeter1981, Goldbeter1984}, function on timescales of at least 30 minutes \cite{Trunnell2011}. We propose that condensate formation provides an alternative mechanism that works in a fraction of the time, though precise measures of timescales will naturally vary between specific sensing systems. 

Finally, we took a step back and asked how well the statistics of the first nucleation event across the subsystems captured the ability to discriminate concentrations: this event can happen on timescales of seconds and offers a route to very fast sensing through phase separation. We observed a high accuracy for fine-tuned supersaturation of $S$, but it was always outperformed by the optimal scenario which exploits the growth of droplets. These findings suggest that nucleation alone can serve as a rapid sensing mechanism, an idea explored in experimental work where the nucleation of large molecular assemblies or phases was used to amplify sensitivity to molecular inputs\cite{Ershova2024-gp, Wintersinger2023-ky}. Related designs based on nucleation in multicomponent systems have been proposed as molecular neural networks capable of distinguishing patterns in the relative concentrations of dozens of molecules\cite{Zhong2017-kc,Evans2024-se}. Together, these results point at the potential for engineered condensate-based systems to perform fast and accurate decision making in synthetic or cellular contexts.

\add{While our model was chosen as a most general picture for biomolecular condensates, an illustrative biological example of our results arises in the context of transcriptional condensates \cite{Hnisz2017}. Phase separation is argued to form droplets mediating interactions between distant genes enabling correlated transcriptional behavior. Above, we concluded that competition for $S$ among subsystems enhances the selectivity of droplet formation. This behavior closely parallels the phenomenon of transcriptional squelching \cite{Gill1988, Natesan1997}, in which the overexpression of a transcription factor leads to excessive recruitment of RNA polymerase, mediator and other transcriptional proteins, thereby indirectly suppressing the expression of non-targeted genes. This agrees with our model, where we see that when $A$ is more abundant than $A’$, we see an amplification in the volume for $A$ droplets (transcriptional hubs for pathway $A$) that form compared to $A’$ due to competition for S (e.g. RNA polymerase).} Transcriptional condensates also compete for binding sites along the DNA. Additionally, constraining our phase separation model by fixing a finite substrate for forming droplets changes the physics described in this paper~\cite{Morin2022-fy,Rouches2025-zo,Strom2024-rr} and may further help to distinguish concentrations in transcriptional regulation by introducing more competition to the system. 

Our key idea of introducing a second system that competes for a shared molecular resource can be tested in \emph{in vitro} systems. Competitive, winner-take-all mechanisms have previously been studied both theoretically~\cite{Kieffer2023-hr,Genot2012-ba} and experimentally~\cite{Cherry2018-qf,Evans2024-se,Chen2024-xq} as a means of sharpening responses in molecular systems. For example, the experiments of \cite{Evans2024-se} involved competing crystalline phases (rather than liquid phases in our system), they showed a higher sensitivity to input concentrations due to the depletion of a shared component (analogous to $S$ in our system). While these prior studies exploit competition for a shared molecular resource through various mechanisms, this work focuses on competition in the context of phase separating droplets and thus guidances for similar \emph{in vitro} experiments with competing liquid phases.

Another interesting extension of our work is to explore how phase separation might support bistable switching. In transcriptional regulation, this type of switch may provide dynamic control over activator–repressor behaviour, reminiscent of genetic toggle switches as engineered in synthetic biology \cite{Gardner2000}. Physically, what sets the timescale for a system initialised with all $A$-droplets to transition to a state dominated by $A'$-droplets through passive phase separation dynamics? In Fig.~\ref{fig:optimal}(c), a system that has equilibrated (so supersaturated concentrations are at the black dots) could then receive an influx of $A$, pushing the dot of the blue curve to the right of the dot of the red curve. After some timescale, droplet nucleation will drive composition of all droplets from one type to the other, where the bistability arises in the sense that there is a nucleation energy barrier to start making the new droplets. Quantifying how these timescales relate to those of the current work and other bistable mechanisms remains an open problem, but could further establish phase separation as a fast, robust and reliable sensing mechanism.

\section*{Acknowledgements}
The study was supported by Agence Nationale de la Recherche grant no
ANR-22-CE95-0005-01 ``DISTANT'' (AMW, TM, HA) and by the CZI Theory Initiative. AM acknowledges support from NIGMS of the National Institutes of Health under award number R35GM151211, National Science Foundation through the Center for Living Systems (grant no. 2317138). We thank Erik Winfree, Eric Dufresne, David Zwicker for discussions.

\onecolumngrid

\appendix

\renewcommand{\thefigure}{S\arabic{figure}}
\setcounter{figure}{0}

\section{Mathematical model for droplet nucleation and growth in ternary mixture}

\subsection{Phase Equilibria}

We consider a ternary mixture of fluids $A$, $S$, and a solvent with volume fractions $\phi_A$, $\phi_S$ and $\phi_H=1-\phi_A-\phi_S$, respectively. We consider the scenario where $A$ can condense and form a dense phase only in conjunction with $S$.
We determine the coexisting phase densities through the Flory-Huggins theory of mixtures, in which the free energy per unit volume for $\bphi=(\phi_A,\phi_S)$ is given by: 
\begin{equation}
\mathcal{F}_{\rm FH}(\bphi)=RTc_0\left[\sum_{i=A,S}\phi_i\ln(\phi_i) + \frac{\phi_H}{v_s}\ln(\phi_H) +\chi\phi_A\phi_S\right],
\end{equation}
where $c_0$ is the total concentration of solutes, and $v_s$ is the (dimensionless) volume ratio between solvent and soluble molecules. We set $v_s=1$ in what follows. The first two terms describe the mixing entropy and the last term describes the interaction energy between the two types of molecules. In what follows, we non-dimensionalize the free energy by rescaling with the characteristic energy density $RT c_0$, where $R$ is the gas constant and $T$ is temperature, thus setting $RT c_0 = 1$ in our equations, so all energies are expressed in units of this natural thermal energy scale. 

For sufficiently negative $\chi$, the system phase separates into co-existing dense and dilute phases, the dense phase being rich in $A$ and $S$ and the dilute phase being rich in solvent. 
We denote the coexisting phase densities by  $\bphi^0 = (\phi_A^0, \phi_S^0)$ and $\bphi^1 = (\phi_A^1, \phi_S^1)$ for the dilute and dense phases, respectively. To determine $\bphi^0$ and $\bphi^1$, we write the free energy density of the two-phase system (ignoring interfacial energies) as 
\begin{equation}
\mathcal{F}_{\rm PS}=\eta\mathcal{F}_{\rm FH}(\bphi^1) + (1-\eta)\mathcal{F}_{\rm FH}(\bphi^0)\end{equation} where we have defined $\eta$ as the dense phase volume fraction. Minimising this with respect to $\eta$, $\bphi^0$ and $\bphi^1$ gives us the phase equilibria. The resulting equations satisfying this optimisation problem are classically interpreted in the following way: the chemical potentials of each component, defined through $\mu_i = {\delta \mathcal{F}_{\rm FH}}/{\delta \phi_i}$ (where $i=A, S$), and the pressure, defined as $P = \mathcal{F}_{\rm FH}(\bphi) - \bphi \cdot \bm{\mu}$, must equate between the two phases. We derive the following expressions for these chemical potentials 
\begin{align}\label{eq:musSI}
    \mu_A(\bphi) = \log \phi_A - \log(1-\phi_A- \phi_S) + \chi \phi_S, \quad \mu_S(\bphi) = \log \phi_S - \log(1-\phi_A- \phi_S) + \chi \phi_A.
\end{align}
and the pressure
\begin{equation}\label{eq:PSI}
    P(\bphi) = -\log(1-\phi_A-\phi_S) + \chi\phi_A\phi_S.
\end{equation}

Equating $\bm{\mu}(\bphi)=[\mu_A, \mu_S]$ and $P$ provides three equations:
\begin{gather}
    \bm{\mu}({\bphi}^0) = \bm{\mu}({\bphi}^1)\\
    \mathcal{F}_{\rm FH}({\bphi}^1)-\mathcal{F}_{\rm FH}({\bphi}^0) = (\bphi^1 - \bphi^0) \cdot \bm{\mu}(\bphi^0). \label{eq:pressuresSI}
\end{gather}
However, there are 4 unknown quantities in $\bphi^0$ and $\bphi^1$. Crucially, the convex hull of the free energy surface is in fact a plane that must be parametrized by a family of tangential chords: this parametrization is done through the relative fractions of $A$ and $S$ in the dense phase: 

\begin{equation}\label{eq:alpha}
    \alpha = \frac{\phi_A^1}{\phi_A^1+\phi_S^1}.
\end{equation}
This will give the desired families of pairs of points, $\bphi^0(\alpha)$ and $\bphi^1(\alpha)$. Finally, we identify another equation satisfied by the phase equilibria and $\eta$ due to the conservation of mass:
\beq\label{eq:dpfraction}
\bphi(t)=(1-\eta) \bphi^0(\alpha)+\eta \bphi^1(\alpha).
\eeq
{where $\phi(t)$ is the supersaturated concentrations of $A$ and $S$ (i.e.\,outside of droplets) at time $t$.
In total, this gives us now 6 equations which can be solved to determine the 6 unknown quantities: $\alpha$, $\eta$ and the coexisting densities. \add{These equations can be solved numerically: the resulting phase diagram is plotted in Fig.\,\ref{fig:sphase-diagram}, including the spinodal decomposition region in blue (where the homogeneous solution is linearly unstable). We compare this to the schematic phase diagram of the main text showing quantitative agreement.} 

In the limit of strong interactions $ -\chi\gg1 $, the dense phase becomes very dense and the dilute phase very dilute. In this limit, the coexisting densities can be calculated analytically from \eqref{eq:musSI}-\eqref{eq:pressuresSI} for a given value of $\alpha$:
\bea
\phi^0_A(\alpha)&=&\alpha e^{\chi(1-\alpha)^2}\label{phi0A}\\
\phi^0_S(\alpha)&=&(1-\alpha) e^{\chi\alpha^2}\label{phi0S}\\
\phi^1_A(\alpha)&=&\alpha (1-e^{\chi\alpha(1-\alpha)})\\
\phi^1_S(\alpha)&=&(1-\alpha) (1-e^{\chi\alpha(1-\alpha)}),
\eea
\add{where we remark here that $\phi^1_A(\alpha)$ and $\phi^1_S(\alpha)$ respect the definition of $\alpha$ as a ratio of compositions.}

So far, we have neglected the role of surface tension. Its introduction results in a modification of the phase equilibria. We denote these modified equilibria by $\hat{\bphi}^0$ and $\hat{\bphi}^1$. To calculate these, we return to the free energy and modify it in the following way: taking the dense phase to be single spherical droplet of radius $R$, we can define the surface area $s(R)=4\pi R^2$, volume $v(R)=4\pi R^3 / 3$ and dense volume fraction $\eta = v/V$ (where $V$ is the system size) to write the new free energy for the single droplet system taking in to account surface tension: 
\begin{equation}
    \mathcal{F}_{\rm PS}=\frac{v}{V}\mathcal{F}_{\rm FH}(\hat\bphi^1) + \left(1-\frac{v}{V}\right)\mathcal{F}_{\rm FH}(\hat\bphi^0) + \frac{\gamma s}{V}
\end{equation}

The modified phase equilibria satisfy a similar set of equations upon balancing chemical potential and pressure across the two phases:
\begin{gather}
    \bm{\mu}(\hat{\bphi}^0) = \bm{\mu}(\hat{\bphi}^1)\\
    \mathcal{F}_{\rm FH}(\hat{\bphi}^1)-\mathcal{F}_{\rm FH}(\hat{\bphi}^0) + \frac{2 \gamma(\alpha)}{R} = (\hat\bphi^1 - \hat\bphi^0) \cdot \bm{\mu}(\hat\bphi^0) \label{eq:pressures_modified}
\end{gather}
where in principle the surface tension depends on the exact composition of the droplet of radius $R$. We assume there is no such dependence in the following. 

Assuming the effect of surface tension is small, the difference between the modified phase equilibria and those derived for $\gamma=0$ above can be derived. We parametrise again by $\alpha = \hat\phi_A^1/(\hat\phi_A^1+\hat\phi_S^1)$ so that we can write $\delta \phi_A^1=\hat\phi_A^1 - \phi_A^1=\epsilon_\gamma\alpha$ and similarly $\delta\phi_S^1=\epsilon_\gamma(1-\alpha).$ After substituting in \eqref{eq:pressures_modified}, we derive \cite{Weber2019}:
\bea
\epsilon_\gamma&=&\frac{2\gamma}{R(\bphi^1-\bphi^0)\cdot(\mathcal{H}(\bphi^1){\bm
    \alpha})}\\
\delta \bphi^1&=&\epsilon_\gamma{\bm\alpha}\\
\delta \bphi^0&=&\epsilon_\gamma
\mathcal{H}(\bphi^0)^{-1}\mathcal{H}(\bphi^1){\bm \alpha}
\eea
where ${\bm\alpha}=(\alpha,1-\alpha)$ and $\mathcal{H}$ is the Hessian of the free energy $\mathcal{F}_{\rm FH}(\bphi)$:
\begin{equation}
    \mathcal{H}(\bphi) = \begin{pmatrix} 
 \frac{{\phi_S}-1}{{\phi_A} ({\phi_A}+{\phi_S}-1)}, & {\chi}-\frac{1}{{\phi_A}+{\phi_S}-1} \\
 {\chi}-\frac{1}{{\phi_A}+{\phi_S}-1}, & \frac{{\phi_A}-1}{{\phi_S} ({\phi_A}+{\phi_S}-1)} \\
\end{pmatrix}
\end{equation}
The direction of $\delta \bphi^0$ is unclear at first glance, but we argue that it must point along the tie line, such that $\delta \bphi^0\propto \bphi^1-\bphi^0.$ This allows us to then define capillary lengths inside and outside the droplet: we define these $\ell_0(\alpha)$ and $\ell_1(\alpha)$ through
\begin{equation}\label{eq:mod_phaseeq}
    \delta \bphi^0 = \frac{\ell_0(\alpha) ( \phi_A^0 + \phi_S^0)}{R}(\bphi^1 -\bphi^0),\quad\delta\bphi^1 = \frac{\ell_1(\alpha)}{R} \bm{\alpha},
\end{equation}
where the dependence on $\alpha$ signifies that, in principle, these lengths depend on the composition of the droplet. We calculate these capillary lengths from the above expressions, writing them in terms of the dilute and dense phase equilibria. The analytic forms of $\epsilon_\gamma$ and $ \mathcal{H}(\bphi)^{-1}$ are not particularly informative. However, in the limit of strong interactions $\bphi_1 \gg \bphi_0$, the expressions for the capillary lengths simplify: $\ell_1\approx 0$, and we have for $\ell_0$:
\begin{equation}
    \frac{\ell_0(\phi_A^0+\phi_S^0)}{R} = \frac{2\gamma (\phi_A^0 + \phi_S^0 + 2\chi \phi_A^0\phi_S^0)}{(1-\phi_A^0 - \phi_S^0)(1+2\chi \phi_A^0\phi_S^0 - \chi^2\phi_A^0\phi_S^0(1-\phi_A^0 - \phi_S^0))R} \approx \frac{2\gamma (\phi_A^0+\phi_S^0)}{R},
\end{equation}
where we have expanded to linear-order in $\phi_A^0$ and $\phi_S^0$ as we assume that they small quantities. Thus, $\ell_0=2\gamma$. We conclude that the derived capillary length in this limit exhibit no dependence on the details of the phase equilibria. 

\begin{figure}
    \centering
    \includegraphics[width=0.9\linewidth]{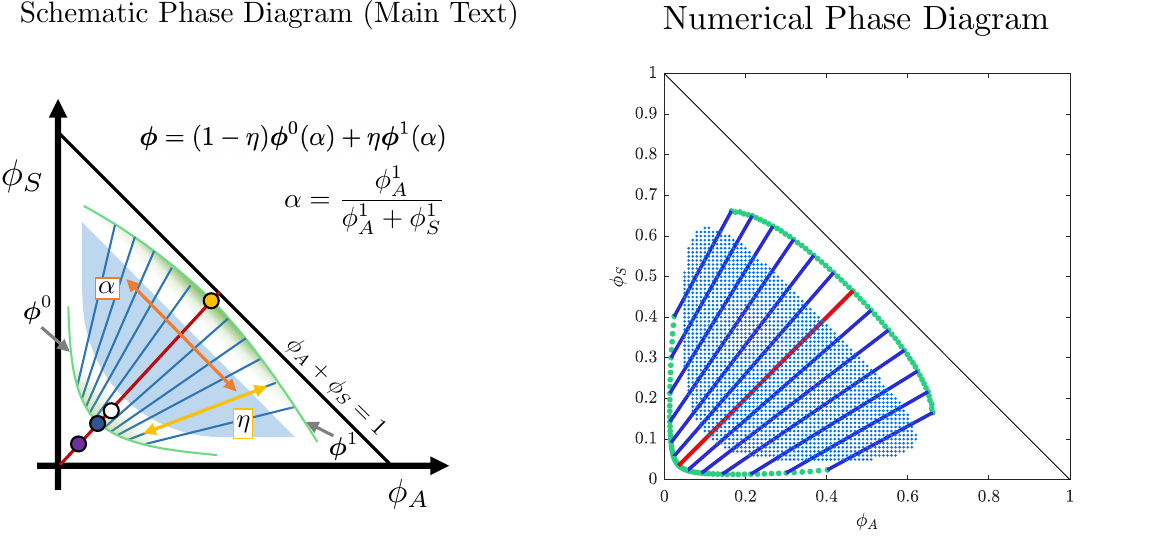}
    \caption{\textit{Comparing schematic and numerical phase diagrams for ternary mixture ---} We numerically solve the $6$ equations to derive the 4 phase equilibria, the composition $\alpha$ and the dense phase volume fraction $\eta.$ We confirm that our schematic phase diagram in the main text faithfully captures the full picture for the model.  }
    \label{fig:sphase-diagram}
\end{figure}

\subsection{Droplet Nucleation}

Having established the phase equilibria for our system, we now turn to the dynamics of droplet formation and growth. Droplet nucleation is driven by thermal fluctuations in the system which drive it from its initial homogeneous (metastable) state to a state with the coexistence of two phases. We demonstrate below that the energy barrier is maximised at a finite $R_c$. Droplets that form with a radius smaller than $R_c$ will not survive: the system will relax back to a homogeneous state. Larger droplets will survive, so we only consider the rate at which droplets of radius $R=R_c$ form as these are the only ones that will persist in the system beyond short times. 

Here we take the following approach: we derive the maximum of the energy barrier which the system needs to overcome in forming a droplet to derive the critical droplet radius. We then use this to establish a kinetic rate for nucleating droplets at this radius through Arrhenius' law (where the rate is set by the exponential of the energy barrier. The main assumption here is that the nucleation rate is dominated by this exponential factor and thus the additional pre-factor can be set independently of the microscopic details. 

To identify the critical radius $R_c(\alpha)$, we write the free energy difference between a (non-equilibrated) droplet volume $v(R)=4 \pi R^3/3$ at concentration $\hat{\bphi}^1$ in a dilute phase with fraction $\bphi'$ and a well-mixed system at concentration $\bphi$, such that $\bphi'=(\bphi - (v/V)\hat{\bphi}^1)/(1-v/V)$: 
\begin{align}
    \Delta F(R)=(V-v(R))\mathcal{F}_{\rm FH}(\bphi')+v(R)\mathcal{F}_{\rm FH}(\hat\bphi^1)+\gamma s(R).
\end{align}
Expanding at small $v$, we find the energy difference $\Delta F(R)$
\begin{align}\label{eq:SI23}
    \Delta {F}(R) = &{v(R)}\left[\mathcal{F}_{\rm FH}(\hat{\bphi}^1) - \mathcal{F}_{\rm FH}(\bphi) - (\hat{\bphi}^1-\bphi)\cdot \bm{\mu}(\bphi)\right] + {\gamma s(R)}.
\end{align}
This difference is maximised at finite $R$: solving $d[\Delta \mathcal{F}]/dR=0$ gives the critical radius 
\begin{equation}\label{eq:SI24}
    R_c = 2\gamma / \left[\mathcal{F}_{\rm FH}(\hat{\bphi}^1) - \mathcal{F}_{\rm FH}(\bphi) - (\hat{\bphi}^1-\bphi)\cdot \bm{\mu}(\bphi)\right].
\end{equation} 
We note the form of this equation is exactly that of \eqref{eq:pressures_modified}. This allows us to derive an equation for the critical droplet radius: the supersaturation $\bphi$ that realises this critical droplet radius $R_c$ is defined through $\bphi=\hat{\bphi}^0(R_c)$.

This gives us two equations for the critical radius: defining $\Delta \phi_A=\phi_A^1 - \phi_A^0$ and $\varepsilon_A = \phi_A - \phi_A^0$ (and similarly for $S$), we derive
\begin{equation}\label{eq:Rc_SM}
    R_c(\alpha) = \frac{\ell_0 ( \phi_A^0 + \phi_S^0)\Delta \phi_A}{\varepsilon_A} = \frac{\ell_0 ( \phi_A^0 + \phi_S^0)\Delta \phi_S}{\varepsilon_S}.
\end{equation}
At first, it appears that the last equality is not necessarily satisfied, but the equality can be seen through \eqref{eq:dpfraction} above which enforces that the dense volume fraction $\eta$ satisfies $\eta = \Delta \phi_A/\varepsilon_A=\Delta \phi_S/\varepsilon_S$.

Substituting the expression from \eqref{eq:SI24} in to \eqref{eq:SI23}, we see the energy barrier is maximised at 
\begin{equation}
    \Delta {F}(R_c) = {v(R_c)}\frac{2 \gamma}{R_c} + {\gamma s(R_c)} = \frac{4 \pi \gamma R_c^2}{3},
\end{equation}
where $R_c$ is given by \eqref{eq:Rc_SM}. From this barrier, we define our nucleation rate for droplets through the exponential energy barrier  \begin{equation}\label{eq:nucrate}
    k_{\rm nuc} = k_0 V\exp\left[-\Delta {F}(R_c)\right] = k_0 V\exp\left[-4\pi\gamma R_c^2 / 3\right] 
\end{equation}
where $k_0$ is assumed to be a constant pre-factor, independent of the compositions or concentrations in the system. 

\subsection{Droplet Growth}

Finally we describe the growth of droplets through diffusive fluxes of material. Suppose that we have a droplet of radius $R_j$ and at composition $\alpha_j$. We note that this $\alpha_j$ may differ from that of the phase equilibria, which we will continue to denote by $\alpha$, or other droplets, say $\alpha_{j'}$. We first define the vector $\bm{J}(R_j, \alpha_j)$ with entries describing the flux of $A$ and $S$ respectively into a droplet of composition $\alpha_j$. We assume this droplet is spherical and that the flux is driven entirely by diffusion, in which case $\bm{J}$ can be written as \cite{Weber2019}:
\beq
{\bm J}(R_j, \alpha_j)=4\pi DR_j(\bphi-\hat\bphi^0(R_j, \alpha_j)),
\eeq
where we have assumed that the concentration right outside the droplet is set by the dilute phase concentration $\hat\bphi^0$ (a gradient forms between that concentration and the supersaturated concentration $\bphi$ far from the droplet, driving the influx of molecules \cite{Weber2019}).
Both the flux $\bm{J}$ and the phase equilibria $\hat\bphi^0$ and $\hat\bphi^1$ depend on the droplet radius $R_j$ and the composition $\alpha_j$, but to ease notation, we neglect this dependence in the following derivation. 

Over the next time step $dt$, the volume of the droplet increases by some amount $dv$. We want to be a bit more careful about how those fluxes drive droplet
growth. As the droplet grows by $dv$, the number of particles $n_A$
and $n_S$ of each
type grows as:
\bea
d{\bm n}=dv \hat\bphi^0+{\bm J} dt,
\eea
where the first term of the right-hand side comes from the fact that
the droplet absorbs particules that were there at concentrations
$\hat\bphi^0$ before. Now we can also write ${\bm n}=\hat\bphi_A^1 v$, so that:
\beq\label{eq:dn}
d{\bm n}=d\hat\bphi^1 v + dv \hat\bphi^1
\eeq
In the case considered so far, $d\hat\bphi^1=0$ because $\hat\phi$ stays on the
same $\alpha$ chord as the droplet grows. But when we'll introduce
competing droplets depleting $S$, the $\alpha$ of the dilute phase and
that of already formed droplets may differ. Recalling $\alpha_j$ the
$\alpha$ of a given droplet, we then have:
\beq
\frac{dv}{dt}(\hat\bphi^1(\alpha_j)-\hat\bphi^0(\alpha_j))+\frac{d\alpha_j}{dt}
\frac{d\hat\bphi^1(\alpha_j)}{d\alpha}={\bm J}=4\pi DR(\hat\bphi-\hat\bphi^0(\alpha_j)),
\eeq
This will remain true even if the $\alpha$ of the dilute phase, $\bphi$ (given by \eqref{eq:alpha}), is different from $\alpha_j$ of the droplet. This system of two
equations can be solved for $dv/dt$ and $d\alpha_j/dt$ to get
expressions for the growth and compositional change of the droplet. 
In the very dense phase limit, $d\hat\phi^1_A/d\alpha=1$ and
$d\hat\phi^1_S/d\alpha=-1$, and we have:
\bea
\frac{dv}{dt}&=&\frac{J_A+J_S}{\hat\phi_A^1-\hat\phi_A^0+\hat\phi_S^1-\hat\phi_S^0}\\
v\frac{d\alpha_j}{dt}&=&J_A-\frac{dv}{dt}(\hat\phi_A^1-\hat\phi_A^0)=\frac{J_A(\hat\phi_S^1-\hat\phi_S^0)-J_S(\hat\phi_A^1-\hat\phi_A^0)}{\hat\phi_A^1-\hat\phi_A^0+\hat\phi_S^1-\hat\phi_S^0}
\eea
or
\bea
\frac{dR_j}{dt}&=&\frac{D}{R_j}\frac{\phi_A-\hat\phi^0_A(\alpha_j)+\phi_S-\hat\phi^0_S(\alpha_j)}{\hat\phi_A^1(\alpha_j)-\hat\phi_A^0(\alpha_j)+\hat\phi_S^1(\alpha_j)-\hat\phi_S^0(\alpha_j)}\label{eq:original_growth},\\
\frac{d\alpha_j}{dt}&=&\frac{3D}{R_j^2}\frac{(\phi_A-\hat\phi^0_A(\alpha_j))(\hat\phi_S^1(\alpha_j)-\hat\phi_S^0(\alpha_j))-(\phi_S-\hat\phi^0_S(\alpha_j))(\hat\phi_A^1(\alpha_j)-\hat\phi_A^0(\alpha_j))}{\hat\phi_A^1(\alpha_j)-\hat\phi_A^0(\alpha_j)+\hat\phi_S^1(\alpha_j)-\hat\phi_S^0(\alpha_j)},
\eea

We now write these dynamics in terms of the phase equilibria $\bphi^0(\alpha_j)$ and $\bphi^1(\alpha_j)$ which in turn makes the dependence on $R_j$ fully explicit. We argued above that $\hat\bphi^1\approx \bphi^1$ and recall \eqref{eq:mod_phaseeq} for $\delta \bphi^0(\alpha_j)=\hat\bphi^0-\bphi^0$ to derive the dynamical equations:
\begin{gather}
    \frac{dR_j}{dt} = \frac{D \ell_0 (\phi_A^0 + \phi_S^0)}{R_j}\left[\frac{1}{R_c(\alpha_j)} - \frac{1}{R_j}\right],\quad R_c(\alpha_j) = \frac{\ell_0 (\phi_A^0 + \phi_S^0)(\Delta\phi_A + \Delta \phi_S)}{\varepsilon_A + \varepsilon_S}\label{eq:dRdt_SM}\\
    \frac{d\alpha_j}{dt} = \frac{3D}{R_j^2} \left[\frac{\Delta\phi_S\varepsilon_A - \Delta\phi_A\varepsilon_S}{\Delta\phi_A + \Delta \phi_S} \right],\label{eq:dadt_SM}
\end{gather}
where we have defined again $\Delta \phi_A(\alpha_j)=\phi_A^1 - \phi_A^0$ and $\varepsilon_A(\alpha_j) = \phi_A - \phi_A^0$ (and similarly for $S$). If $\alpha_j$ matches that of the dilute phase, $\alpha$, then the right-hand side of \eqref{eq:dadt_SM} vanishes, as argued under \eqref{eq:Rc_SM}. Otherwise, the dense phase composition $\alpha_j$ is not the one set by the supersaturated concentrations $\bphi$, thus the conservation of mass constraint does not apply so the right-hand side can be non-zero. It is straightforward to check that the definition for $R_c$ derived here through equating the effects of diffusive fluxes in to the droplet and surface tension is identical to the two equations derived from the energy barrier for nucleation. 

We now have a complete description of droplet nucleation and growth in a ternary mixture. We remark that the dynamics for nucleation and growth at time $t$ are set by the (non-equilibriated) concentrations $\bphi(t)$. In what follows, we will consider a system comprised of two ternary mixtures that each evolve under the dynamics just described. We quantify the different phase separation dynamics across these two mixtures when changing the concentration of $A$ in one system while keeping the other fixed. We will study the effect of interactions between these subsystems, namely $(i)$ exchanging of the building block $S$ and $(ii)$ the replenishing of $A$ in the dilute phase upon forming droplets. We conclude on how to optimise the sensitivity of this selection process based on relative concentrations of $A$. 

\subsection{Numerical procedure for simulations of a ternary mixture}

 \begin{enumerate}
\item We initialise the system with volume fractions of $\phi_A$ for $A$ and $\phi_S$ for $S$ molecules and say the system is homogeneous, thus without any droplets. This implies the total number $A_{\rm free} = V\phi_A$ of free $A$ molecules and $S_{\rm free} = V\phi_S$ of free $S$ molecules. We define also the free volume as the volume not contained in droplets, $V_{\rm free}$.
 \item From the volumes fractions, we can compute the 4 phase equilibria ($\bphi^0$ and $\bphi^1$) plus the composition $\alpha$ and dense volume fraction $\eta$.
   \item Create a new droplet of composition $\bphi^1$ and radius $R_c$ defined in \eqref{eq:Rc_SM} 
 with rate $k_{\rm nuc}\times dt$ as given by \eqref{eq:nucrate} for volume $V_{\rm free}$.
 \item Use conservation of mass to update
   $\phi_A$ and $\phi_S$ after droplet creation: $A_{\rm
     free}\leftarrow A_{\rm free}-(\phi^1_A-\phi_A) (4\pi/3)R_c^3$, $S_{\rm
     free}\leftarrow S_{\rm free}-(\phi^1_A-\phi_S) (4\pi/3)R_{c}^3$. Also update free volume $V_{\rm free}\leftarrow V_{\rm free}-(4\pi/3)R_c^3$. 
   \item Grow each extant droplet, and update their composition (which is
     different for each droplet) using \eqref{eq:dRdt_SM},\eqref{eq:dadt_SM}. (We solve these equations with an explicit Euler method for 1 time-step). If a droplet $j$ at composition $\alpha_j$ has radius $R_j$ below 10\% of the critical droplet size, $0.1R_c(\alpha_j,t)$, we classify it as a droplet that is shrinking and thus dissolve it.
\item \add{We use conservation of mass to update $\phi_A$ and $\phi_S$ following the increments of size and composition of each droplet. We find the number of free $A$ and $S$ molecules by taking the difference between the total molecule number (e.g. $A_{\rm total}$) and the molecule number in droplets (e.g. $A_{\rm drop} = \sum_j R_j^3 \alpha_j$). We also calculate the free volume, namely the volume of the system not taken up by droplets, as $V_{\rm free} = V_{\rm tot} - \sum_j R_j^3$. The dilute volume fractions are then calculated through $phi_A = A_{\rm free}/V_{\rm free}$ and similarly for $\phi_S$. If $S$ is exchanged between two subsystems, the $\phi_S$ in each subsystem is replaced by the average of the two values, which in turn leads to new values for $S_{\rm total}$ in each subsystem: we calculate this here from summing over the $S$ molecules in droplets (e.g. $S_{\rm drop} = \sum_j R_j^3 (1-\alpha_j)$) and outside for this new $\phi_S$ ($S_{\rm free}=\phi_S*V_{\rm free}$).  In the case where $A$ and $A'$ are replenished, then the volume fractions $\phi_A$ and $\phi_{A'}$ are held constant through the simulations and the steps detailed here do not apply for $A$ molecules.}
  \item Advance time by a time-step $dt$. Repeat from 2.\,until total time reaches $T_f$. 
\end{enumerate}

 There are simulation results given in the main text where we consider 2 ternary mixtures which exchange $S$. The procedure given here is modified trivially to account for these extra dynamics: after updating the volume fractions in 6.\,, we compute the total volume fraction of $S$ in the dilute phase across the two systems as $\phi_S$ = $(S_{\rm free}(\bphi)+S_{\rm free}(\bphi')) / (V_{\rm free}(\bphi) + V_{\rm free}(\bphi')).$ By equating the volume fraction of $S$ outside of droplets across the two subsystems, we model an effective diffusive flux of $S$: from the subsystem with more $S$ to the subsystem with less.

\subsection{Measuring Accuracy}

In the current work, we consider two sensing mechanism: \textit{first-nucleation} sensing and \textit{droplet proportion} sensing. In \textit{first-nucleation} sensing, we say that the cell infers whether $A$ is more concentrated than $A'$ depending on whether $A$ nucleates the first droplet (i.e.\,before $A'$). If $A'$ nucleates first, the cell infers that $A'$ is more concentrated. If neither subsystem nucleates a droplet before time $T$, we say that the cell guesses randomly with a $50\%$ chance of concluding $A$ or $A'$ is more concentrated. Over many simulations at different concentrations of $\phi_A$ (keeping $\phi_{A'}=0.025$ and the decision time $T_f$ constant), we measure the probability that the cell infers that $A$ is more concentrated than $A'$, denoting this outcome by $\mathcal{A}$ and defining the probability $P(\mathcal{A})$. Given that we have fixed $\phi_{A'}$ and $T_f$, this is a function solely of $\phi_A-\phi_{A'}$ that we measure numerically: $P(\mathcal{A})=f_{\rm measured}(\phi_A - \phi_{A'})$. This is plotted in the main text in Figures 3b, 4b and 5b. 

Perfect sensing would be realized by a Heaviside step function: \begin{equation}
    f_{\rm perfect}(\phi_A-\phi_{A'}) = \begin{cases}
        0  & \phi_A<\phi_{A'} \\
        1  & \phi_A>\phi_{A'} \\
    \end{cases}
\end{equation}
First defining $\psi=\phi_A-\phi_{A'}$, we then turn to quantifying the accuracy of the first sensing mechanism. We use the following definition:
\begin{equation}
    {\rm Accuracy} = 1 - \int_{-\infty}^{\infty} d\psi \:|f_{\rm perfect}(\psi) - f_{\rm measured}(\psi)|\frac{1}{\sqrt{2\pi\sigma^2}}\exp\left(-\frac{\psi^2}{2\sigma^2}\right),\quad  \sigma=0.02\times \phi_{A'}.
\end{equation}
\add{The first term in the integrand is the absolute difference between the measure and perfect function. The second term is a Gaussian function centered at $\psi=0$ with standard deviation which is $2\%$ of $\phi_{A'}$. The choice of a Gaussian function is arbitrary: we choose it such that the accuracy measures with what probability first-nucleation sensing infers whether $A$ or $A'$ is more concentrated when the difference in concentrations is $<5\%$, with greater reward for distinguishing smaller concentrations.}

In \textit{droplet-proportion} sensing, we say that the cell infers whether $A$ is more concentrated than $A'$ depending on the volume of $A$ droplets as a fraction of the total volume of droplets. As we did above for {first-nucleation} sensing, we say that if no droplets nucleate, the cell guesses with $50\%$ chance that it is $A$ that is more concentrated. 

As above, $T_f$ and $\phi_{A'}$ are fixed, so we can again extract $f_{\rm measured}(\psi)$ from numerical simulations and measure the accuracy in the same manner. Plots of the measured accuracy across different set-ups are given in the main text and here in the supplementary material in Fig.~\ref{sfig:Tf}.

\section{Different decision making times $T_f$}

Here we contrast the numerical results for the accuracy in the 4 set-ups we have considered when we vary the decision time, $T_f$ from 30 seconds to 10 minutes. The results are given in Figure \ref{sfig:Tf}. For $T_f=$30 seconds, the accuracy curves look quite similar. This is because the volumes of the dense phases are dominated by nucleation events initially, as there is less time for droplet growth. Indeed, it is only as material is depleted that we should expect to see any effect of the competition for $S$ or replenishing of $A$. It is also why the black curve for the first nucleation event is fairly constant across the three time points: it is agnostic to diffusive fluxes between the subsystems and set by the initial nucleation rates which are constant across the three time points. 

The difference between the systems is thus understandable more noticable for $T_f=$2 minutes. For smaller $\varepsilon_{S'}$ the accuracy is controlled by the first few nucleation events, hence why the curves are very similar. However, at large $\varepsilon_{S'}$ the curves are very different: here, many droplets are forming and growing. While the orange and yellow curves are lower than at $T_f=$30 seconds, the green curve is higher. We argue that this is because growth in the orange and yellow systems allows both subsystems to develop droplets, meaning a more even composition of the dense phase. For the green system, where $A$ is replenished, diffusive fluxes are leading to antagonistic interactions between the two subsystems, driving the dissolution of droplets in one system through diffusive fluxes of $S$ between the subsystems. This effect becomes greater at larger times: for $T_f=$10 minutes, almost all of the dense phase is made up of one type of droplet. While this homogeneous composition of the dense phase is expected in the (long-time) equilibrium state, we show here that it can be realised on finite, relevant timescales for a sufficient supersaturation of $S$. 

\begin{figure}
    \centering
    \includegraphics[width=0.9\linewidth]{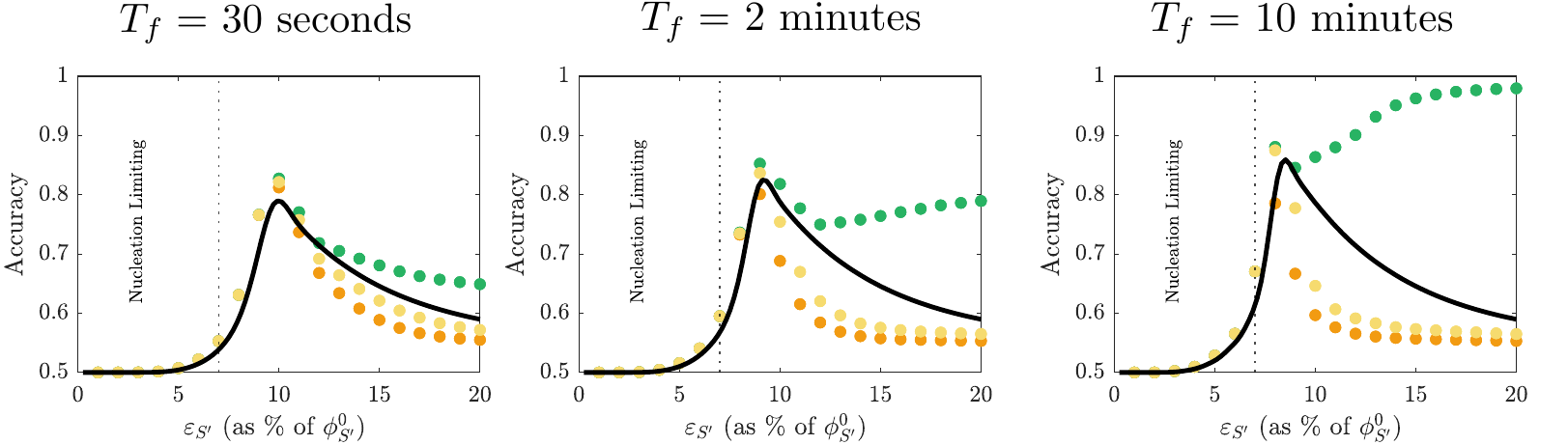}
    \caption{\textit{Accuracy for different decision making times $T_f$ ---} We compare the accuracy of the 4 different sensing set-ups in our system for different decision making times. The orange curve describe two independent ternary mixtures, whereas the yellow points are for two mixtures which exchange $S$. The green curve describes two mixtures that exchange $S$ and replenish $A$ upon forming droplets. The black curve is the accuracy for first-nucleation sensing, so details of $S$ exchange of $A$ replenishing do not play a role. }
    \label{sfig:Tf}
  \end{figure}

\end{document}